\documentclass[10pt]{article}
\usepackage{amssymb}
\usepackage{amscd}
\usepackage{latexsym}
\usepackage{boldtensors}
\usepackage{amsmath}
\usepackage{amsthm}
\usepackage{graphicx}
\newcommand{\newsec}{\setcounter{equation}{0}\section}%
\newcommand{\Zz}{{\mathbb Z}}
\newcommand{\Nn}{{\mathbb N}}
\newcommand{\Rr}{{\mathbb R}}

\newcommand{\Ee}{{\mathbb E}}
\newcommand{\Pp}{{\mathbb P}}
\def\be {\begin{equation}}
\def\ee{\end{equation}}
\def\bea{\begin{eqnarray}}
\def\eea{\end{eqnarray}}
\def\Tr{{\rm \,Tr\,}}

\def\d{{\rm d}}
\def\i{{\rm  i}}

\def\supp{{\rm supp\,}}
\newtheorem*{theorem}{Theorem}
\newtheorem{lemma}{Lemma}[section]

\newtheorem{proposition}{Proposition}[section]
\newtheorem*{corollary}{Corollary}

\theoremstyle{remark}  \newtheorem{remark}{Remark}

\begin{document}
\title{
\large\bf Simultaneous occurrence of off-diagonal long-range order and infinite permutation cycles in systems of interacting atoms\footnote{Work supported by OTKA Grant No. K128989.}}
\author{Andr\'as S\"ut\H o\\Wigner Research Centre for Physics\\P. O. B. 49, H-1525 Budapest, Hungary\\
E-mail: suto.andras@wigner.hu\\}
\date{}
\maketitle
\thispagestyle{empty}
\begin{abstract}
\noindent
Based on the paper "Fourier formula for quantum partition functions", arXiv:2106.10032 [math-ph], we show that in an infinite system of identical bosons interacting via a positive-type pair potential there is off-diagonal long-range order if and only if a nonzero fraction of the particles form infinite permutation cycles. In particular, there is Bose-Einstein condensation if and only if the diverging cycle lengths increase at least as fast with $N$, the number of particles, as $N^{2/d}$ in $d\geq 3$ dimensions. This extends a similar result known for the ideal Bose gas.
\end{abstract}
\newsec{Introduction}

In 1953 Feynman made an attempt to prove the superfluid transition in liquid helium from first principles [F1].
He applied the Feynman-Kac formula [K1,2], the imaginary-time version of his path-integral method [F2], used the analogy with Bose-Einstein condensation (BEC) in the ideal Bose gas and argued that the transition must be signalled by the appearance of long permutation cycles. A proof for the ideal Bose gas, showing that "long" actually means macroscopic cycles (each containing a nonzero fraction of all the particles), was given only half a century later [S2], and its extension to interacting systems has not been achieved since then. Based on a preceding paper [S3] here we present the proof for identical bosons interacting via positive-type pair potentials. The statement itself is meaningful only if one applies the first-quantized method when symmetrization is done by an explicit summation over the permutations; in the algebraic, second-quantized description the problem does not even arise. This forces to pose the question about the physical meaning of the permutation cycles. A possible interpretation will be presented in a subsequent publication.

It is precisely because of the predominant application of second quantization that Feynman's idea had been in a winter sleep for decades. While functional integration offers an elegant and efficient approach to many problems, in quantum statistical physics it proves to be a rather heavy method: to see this it suffices to have a look at Ginibre's formidable work [G1-4] on quantum gases at low fugacity. A substantial reappearance of path integrals in this field took place in the early nineties. A paper by Aizenman and Lieb [AL] applied it to prove the partial survival of Nagaoka ferromagnetism in the Hubbard model at positive temperatures. T\'oth [T] proved BEC of hard-core bosons on the complete graph. Aizenman and Nachtergaele [AN] studied ordering in the ground state of quantum spin chains. Ceperly [Ce] applied the path-integral Monte Carlo method to a thorough numerical analysis of the superfluidity of liquid helium. The present author picked up the thread left by Feynman and discussed BEC of particles in continuous space in connection with the probability distribution of permutation cycles [S1]. In retrospect, the most interesting finding of that paper was that for spinning bosons the Bose statistics in itself induces ferromagnetic correlations independently of any other interaction among the particles, and the zero-field magnetic susceptibility is proportional to the expected length of the cycle that contains a given particle. This latter means that a ferromagnetic long-range order is simultaneous with the appearance of macroscopic permutation cycles. Another result in [S1] connects already to our actual concern. It was shown that BEC implies infinite permutation cycles in the ideal Bose gas. The implication in the opposite sense was proved only ten years later [S2] with some interesting details that we shall recall here together with a simplified proof. The revival of interest in the relation between BEC and infinite cycles gave rise to many other papers, e.g. [BSch], [Sch], [U1], [U2], [BCMP], [DMP], [ACK]. Besides, there appeared a new field of research on Hamiltonian models of random permutations, apparently more amenable to study by functional integration and large deviations analysis; see e.g. [BU1], BU2], [BUV], [EP], [AD].

In this paper we deal with the problem of interacting particles. A formal definition of the quantities appearing below will be given in Section 2.

\begin{theorem}
Consider $N$ identical bosons on a $d\geq 3$-torus of side $L$ at inverse temperature $\beta$ that interact via a pair potential $u:\Rr^d\to\Rr$ of the following properties:
\begin{itemize}
\item[(i)]
$u$ is of positive type with a sufficiently fast decaying Fourier transform $\hat{u}(\geq 0)$.
\item[(ii)]\label{decay}
$u(x)=O\left(|x|^{-d-\eta}\right)$ with some $\eta>0$ as $x\to\infty$ \hspace{5pt} (condition for periodization).
\end{itemize}

\noindent
With the notations
\begin{itemize}
\item[--]
$\lambda_\beta\sim\sqrt{\beta}$ the thermal wave length,
\item[--]
$\rho=N/L^d$
\item[--]
$\rho^{N,L}_n$ the density of particles in permutation cycles of length $n\geq 1$,
\item[--]
$\rho^{N,L}_0$ the density of zero-momentum particles,
\item[--]
$\rho_n=\lim_{N,L\to\infty,N/L^d=\rho}\,\rho^{N,L}_n$, $n\geq 0$,
\item[--]
$\sigma^{N,L}_1$ the one-particle reduced density matrix,
\item[--]
$\langle x|\sigma^{N,L}_1|0\rangle$ the integral kernel of $\sigma^{N,L}_1$,
\item[--]
$\sigma_1(x)=\lim_{N,L\to\infty,N/L^d=\rho}\,\langle x|\sigma^{N,L}_1|0\rangle$
\end{itemize}
we have the following.

\noindent
\begin{enumerate}
\item
\be\label{cond1}
\rho-\sum_{n=1}^\infty \rho_n
\ \leq\
\sigma_1(x)
\ \leq
\rho-\sum_{n=1}^\infty \rho_n +\sum_{n=1}^\infty \rho_n \exp\left\{-\frac{\pi x^2}{n\lambda_\beta^2}\right\}
\ee
implying
\be\label{lim-cond1}
\lim_{x\to\infty}\sigma_1(x)= \rho-\sum_{n=1}^\infty \rho_n.
\ee
\item
For any $c>0$
\be\label{cond2}
\rho^\infty_c
\ \leq\ \rho_0
\ \leq\ \lim_{c'\downarrow 0} \rho^\infty_{c'}.
\ee
Here
\be\label{rho-infty-c}
\rho^\infty_c= \lim_{N,L\to\infty, N/L^d=\rho}
\sum_{n= \lfloor cN^{2/d}\rfloor}^N\rho^{N,L}_n\int_{\Rr^d} \frac{\nu^{N,L}_n(y)\,\d y}{\sum_{z\in\Zz^d}\exp\left\{-\frac{\pi n \lambda_\beta^2}{L^2}z\cdot\left(z+2Ly\right)\right\}}.
\ee
$\int\nu^{N,L}_n\d y=1$, $\nu^{N,L}_n$ is concentrated to an $O(1/\sqrt{n})$ neighborhood of the origin and tends to Dirac's $\delta_0$, making sure that $\rho_0>0$ if
\be\label{cond-BEC}
\lim_{N,L\to\infty, N/L^d=\rho}\sum_{n\geq cN^{2/d}}\rho^{N,L}_n>0.
\ee
If the infinite cycles are exclusively macroscopic then
\be\label{rho0-macr-cycles}
\rho_0=\lim_{\varepsilon\downarrow 0}\,\lim_{N,L\to\infty, N/L^d=\rho}\sum_{n= \lfloor \varepsilon N\rfloor}^N\rho^{N,L}_n.
\ee
\end{enumerate}
\end{theorem}

\begin{remark}
We do not lose much of the generality by supposing that $u(x)$ is central, i.e. depends only on $|x|$. However, we need only $u(-x)=u(x)$, a property of real positive-type functions, and its consequence $\hat{u}(-x)=\hat{u}(x)$. The sufficiently fast decay for $\hat{u}$ means at least $\int \hat{u}(x) x^2 \d x<\infty$.
\end{remark}

\begin{remark}
$\sum_{n=1}^\infty \rho_n$ is the density of particles in finite cycles in the infinite system. Thus, (\ref{lim-cond1}) tells us that for positive-type pair potentials there is off-diagonal long-range order (ODLRO) if and only if there exist infinite cycles with nonzero probability according to the infinite-volume Gibbs measure. The upper bound is sharp also at $x=0$ because $\sigma_1(0)=\rho$, and it holds with equality for all $x$ if there is no interaction  [U2, Theorem 1].
\end{remark}

\begin{remark}
The upper bound in (\ref{cond2}) shows that for BEC cycles whose length diverges at least as fast as $N^{2/d}$ are necessary. The same condition was found for the ideal Bose gas.
If $n\geq cN^{2/d}$ then $1/\sqrt{n}\leq 1/(\sqrt{c}\rho^{1/d}L)$. Therefore,
$L|y|=O(1)$ in the domain of concentration of $\nu^{N,L}_n(y)$, and
the sum over $\Zz^d$ in (\ref{rho-infty-c}) remains bounded as $N,L\to\infty$. Within the terms belonging to $n\propto N$, $L|y|=O(N^{1/d-1/2})$, and the whole integral tends to 1, which explains Eq.~(\ref{rho0-macr-cycles}).
Given the above theorem, the proof of BEC consists in showing that (\ref{cond-BEC}) holds true for $\rho\lambda_\beta^d$ large enough. This is the subject of a forthcoming paper.
\end{remark}

In Section 2 we survey the basic formulas we need for the sequel. They arise from our preceding paper [S3] and the reader is referred to that work for details. In Section 3 we summarize our earlier findings for the ideal Bose gas. Although the theorem covers this case, we present a simple new proof specific for the noninteracting gas, and recall without proof some more results from [S2] that we do not obtain here: in case of BEC there is a countable infinite number of infinite cycles with a known size distribution, each macroscopic and together containing the totality of the condensate and nothing else. Section 3 is ended with a description of the limit shape of partitions of $N$ for the ideal Bose gas. To facilitate the comparison we use Vershik's [V] definitions and notations. These results are also taken over from [S2] where they were not made explicit. Their particularity is that the canonical Gibbs measure is not multiplicative, and the grand-canonical Gibbs measure cannot be used to describe the limit shape composed of the macroscopic elements of the partition. Finally, the proof of the theorem is given in Section 4.

\newsec{Key formulas}\label{key}

Our formulas are valid for pair potentials whose Fourier transform $\hat u$ exists and $\hat u\in L^1\cap C(\Rr^d)$.
From [S3] we recall the expression of the canonical partition function for $N$ particles on a $d\geq 1$-torus $\Lambda$ of side $L$:
\be\label{QNL}
Q_{N,L}=\frac{1}{N}\sum_{n=1}^N G^N_n
= \frac{1}{N}\sum_{n=1}^{N-1}\left(\sum_{p=1}^{N-n}\frac{1}{p!}\sum_{n_1,\dots,n_p\geq 1:\sum_1^p n_l=N-n}
\frac{G\left[n,\{n_l\}_1^p\right]}{\prod_1^p n_l}\right)+\frac{1}{N}G[N].
\ee
For $G\left[n,\{n_l\}_1^p\right]\equiv G\left[\{n_l\}_0^p\right]$ ($n_0=n$) see [S3], to be compared with Eqs.~(\ref{F(x)})-(\ref{F(x)bis}) below.
The only deviation from [S3] is that now the elements of the partition which are lengths of permutation cycles are numbered from 0 to $p$ and the zeroth cycle is treated separately: it is the cycle that contains 1. Still, $G\left[\{n_l\}_0^p\right]= G\left[\{n_{\pi(l)}\}_0^p\right]$ for any permutation $\pi$ of $0,1,\dots,p$. Because
\[
\sum_{p=1}^{N-n}\frac{1}{p!}\sum_{n_1,\dots,n_p\geq 1:\sum_1^p n_l=N-n}\frac{1}{\prod_1^p n_l}=1\quad (n=1,\dots,N-1),
\]
the expression in parentheses in Eq.~(\ref{QNL}) denoted by $G^N_n$
is the average of $G\left[n,\{n_l\}_1^p\right]$ over the partitions of $N-n$.
The $n=N$ term is $G^N_N\equiv G[N]$, all the particles form a single cycle.
We define a density $\rho^{N,L}_n$ ($1\leq n \leq N$) by the equation
\be\label{rhoNLn/rho}
\frac{\rho^{N,L}_n}{\rho}:=\frac{G^N_n}{NQ_{N,L}}
=: P_{N,L}(\xi_1=n)
\ee
where $\rho=N/L^d=\sum_{n=1}^N\rho^{N,L}_n$ is the total density of particles. The middle member of Eq.~(\ref{rhoNLn/rho}) can be interpreted either as the expected value of the fraction of particles in $n$-cycles or as the probability that particle no.1 is in a cycle of length $\xi_1=n$, both according to the canonical Gibbs distribution. In [S1] and [S2] $P_{N,L}(\xi_1=n)$ was the central object.

The results of this paper will be obtained by analyzing
\bea\label{F(x)}
F\left[n,\{n_l\}_1^p\right](x)=
 \sum_{\alpha^2_1,\alpha^3_1,\alpha^3_2,\dots,\alpha^N_{N-1}=0}^\infty\
\prod_{1\leq j<k\leq N}\frac{\left(-\beta\right)^{\alpha^k_j}}{\alpha^k_j !} \prod_{r=1}^{\alpha^k_j} \frac{1}{L^d} \sum_{z^k_{j,r}\in\Zz^d\setminus\{0\}}\hat{u}\left(\frac{z^k_{j,r}}{L}\right) \int_0^1\d t^k_{j,r}
\nonumber\\
\left[\prod_{l=1}^p \delta_{Z^l_1,0}
\exp\left\{-\frac{\pi n_l \lambda_\beta^2}{L^2}\left[\overline{\left(Z^l_{^\cdot}\right)^2}-\overline{Z^l_{^\cdot}}^2\right]\right\}
\sum_{z\in\Zz^d}\exp\left\{-\frac{\pi n_l \lambda_\beta^2}{L^2}\left(z+\overline{Z^l_{^\cdot}}\right)^2\right\}
\right.
\nonumber\\
\left.
\delta_{Z^0_1,0}\exp\left\{-\frac{\pi n \lambda_\beta^2}{L^2}\left[\overline{\left(Z^0_{^\cdot}\right)^2}-\overline{Z^0_{^\cdot}}^2\right]\right\}f_n\left(x;\overline{Z^0_{^\cdot}}\right)\right]
\nonumber\\
\eea
where
\bea\label{fnx}
\lefteqn{f_n\left(x;\overline{Z^0_{^\cdot}}\right)=\frac{1}{L^d}\sum_{z\in\Zz^d}\exp\left\{-\frac{\pi n\lambda_\beta^2}{L^2}(z+\overline{Z^0_{^\cdot}})^2\right\} \cos \left[\frac{2\pi}{L}z\cdot x\right]     }
\nonumber\\
&&=\frac{1}{\lambda_{n\beta}^d}\sum_{z\in\Zz^d}\exp\left\{-\frac{\pi(x+Lz)^2}{n\lambda_\beta^2}\right\}
\cos\left[\frac{2\pi}{L}\overline{Z^0_{^\cdot}} \cdot (x+Lz)\right].
\eea
For $x=0$ the zeroth cycle adds a similar contribution to $L^d F\left[n,\{n_l\}_1^p\right](0)$ as all the other cycles, giving a hint to the relation
\be\label{F(0)}
G\left[\{n_l\}_0^p\right]=e^{-\beta \hat{u}(0)N(N-1)/(2L^d)}L^d F\left[n_0,\{n_l\}_1^p\right](0).
\ee
In (\ref{F(x)}) there is a summation with respect to $\alpha^k_j$ for every pair $j<k$, and what is in the outmost square brackets is under all the summations/integrals. From [S3] we recall that $\overline{Z^l_{^\cdot}}$ and $\overline{\left(Z^l_{^\cdot}\right)^2}$ are averages of $Z_q(t)$ and $Z_q(t)^2$, respectively: if
\be
N_l=\sum_{l'=0}^l n_{l'}\quad (l=0,1,\dots,p),\quad N_0= n_0= n, \quad N_p=N
\ee
then for $q\in C_l:=\{N_{l-1}+1,\dots,N_l\}$
\bea\label{Zqt}
Z_q(t)=-\sum_{j=1}^{q-1}\sum_{k=q}^{N_l}\sum_{r=1}^{\alpha^{k}_{j}}{\bf 1}\{  t^{k}_{j,r}\geq t\} z^{k}_{j,r} +\sum_{j=q}^{N_l}\sum_{k=N_l+1}^{N}\sum_{r=1}^{\alpha^{k}_{j}}{\bf 1}\{  t^{k}_{j,r}\geq t\} z^{k}_{j,r}\nonumber\\
-\sum_{j=1}^{q}\sum_{k=q+1}^{N_l}\sum_{r=1}^{\alpha^{k}_{j}}{\bf 1}\{  t^{k}_{j,r}<t\} z^{k}_{j,r} +\sum_{j=q+1}^{N_l}\sum_{k=N_l+1}^{N}\sum_{r=1}^{\alpha^{k}_{j}}{\bf 1}\{  t^{k}_{j,r}<t\} z^{k}_{j,r}.
\eea
In particular,
\be\label{Z^l_1}
Z^l_1\equiv Z_{N_{l-1}+1}(0)
=-\sum_{j=1}^{N_{l-1}}\sum_{k\in C_l}\sum_{r=1}^{\alpha^{k}_{j}}z^{k}_{j,r}+\sum_{j\in C_l} \sum_{k=N_l+1}^N \sum_{r=1}^{\alpha^{k}_{j}}z^{k}_{j,r}.
\ee
We have
\be
\overline{Z^l_{^\cdot}}
=\frac{1}{n_l}\sum_{q\in C_l}\int_0^1 Z_q(t)\d t,\quad \overline{\left(Z^l_{^\cdot}\right)^2}=\frac{1}{n_l}\sum_{q\in C_l}\int_0^1 Z_q(t)^2\d t,
\ee
therefore $\overline{\left(Z^l_{^\cdot}\right)^2}-\overline{Z^l_{^\cdot}}^2$ is nonnegative; actually it is positive unless $Z_q(t)= 0$ a.s. which occurs if in (\ref{Zqt}) all
$\alpha^k_j=0$ [S3,~Remark~5].
The explicit form of $\overline{Z^l_{^\cdot}}$ is given by
\bea\label{avZ^l}
n_l\overline{Z^l_{^\cdot}}
=-\sum_{\{j<k\}\subset C_l}(k-j)\sum_{r=1}^{\alpha^k_j}z^k_{j,r}
- \sum_{l'=1}^{l-1}\sum_{j\in C_{l'}}\sum_{k\in C_l}\sum_{r=1}^{\alpha^k_j}(k-N_{l-1}-1+t^k_{j,r})z^k_{j,r}
\nonumber\\
+\sum_{l'=l+1}^p\sum_{j\in C_l}\sum_{k\in C_{l'}}\sum_{r=1}^{\alpha^k_j}(j-N_{l-1}-1+t^k_{j,r})z^k_{j,r}.
\eea
For $l=0$ the general formulas simplify to
\be\label{Z01}
Z^0_1=\sum_{j=1}^n\sum_{k=n+1}^N\sum_{r=1}^{\alpha^k_j}z^k_{j,r}\ ,
\ee
\be\label{avZ-0}
n\overline{Z^0_{^\cdot}}
=-\sum_{\{j<k\}\subset C_0}(k-j)\sum_{r=1}^{\alpha^{k}_{j}}z^{k}_{j,r}
+\sum_{j\in C_0}\sum_{k=n+1}^N \sum_{r=1}^{\alpha^{k}_{j}}(j-1+t^{k}_{j,r})z^{k}_{j,r}\ .
\ee
The more complicated expression for $\overline{\left(Z^0_{^\cdot}\right)^2}$ will be given later.

Although it is not obvious from Eq.~(\ref{F(x)}),  $F\left[n,\{n_l\}_1^p\right](x)$ is positive. This is because
\bea\label{F(x)bis}
e^{-\beta \hat{u}(0)N(N-1)/(2L^d)}F\left[n_0,\{n_l\}_1^p\right](x)=\hspace{8cm}
\nonumber\\
=\int W_{0x}^{n_0\beta}(\d\omega_0) \prod_{0\leq j<k\leq n_0-1} \exp\left\{-\int_0^\beta u_L(\omega_0(k\beta+t)-\omega_0(j\beta+t)) \d t\right\}
\nonumber\\
\int_\Lambda \d x_1 \int W_{x_1x_1}^{n_1\beta}(\d\omega_1) \prod_{0\leq j<k\leq n_1-1} \exp\left\{-\int_0^\beta u_L(\omega_1(k\beta+t)-\omega_1(j\beta+t))
 \d t\right\}
\nonumber\\
 \cdots \int_\Lambda \d x_p \int W_{x_px_p}^{n_p\beta}(\d\omega_p) \prod_{0\leq j<k\leq n_p-1} \exp\left\{-\int_0^\beta u_L(\omega_p(k\beta+t)-\omega_p(j\beta+t)) \d t\right\}
\nonumber\\
\prod_{0\leq l'<l\leq p}\prod_{j=0}^{n_{l'}-1}\prod_{k=0}^{n_l-1} \exp\left\{-\int_0^\beta u_L(\omega_l(k\beta+t)-\omega_{l'}(j\beta+t)) \d t\right\}
\eea
where $W^\beta_{yy'}(\d\omega)$ is the Wiener measure on the torus $\Lambda$ for trajectories that start in $y$ at time 0 and end in $y'$ at time $\beta$, and
\be
u_L(y)=\sum_{z\in\Zz^d} u(y+Lz).
\ee
The derivation of the equality between (\ref{F(x)}) and (\ref{F(x)bis}) was the subject of the paper [S3] in the case of $x=0$. When we computed the partition function, $W_{0x}^{n_0\beta}(\d\omega_0)$ was replaced by $W_{xx}^{n_0\beta}(\d\omega_0)$ and integrated over $x$ as well. (Even though we could use $\int_\Lambda\d x\int W_{xx}^{n_0\beta}(\d\omega_0)\cdots=L^d\int W_{00}^{n_0\beta}(\d\omega_0)\cdots$ due to translation invariance).
The modification to be done for a fixed $x\neq 0$ can be deduced from the procedure we followed there. Fourier expansion of the entries of the stochastic integral makes $x$, as all the other spatial variables, appear in the argument of a complex unit. However, while the complex units belonging to integration variables disappear via integration over the torus, that of $x$ remains and, because $F\left[n_0,\{n_l\}_1^p\right](x)$ is real, we can take its real part: that is the cosine figuring in $f_n\left(x;\overline{Z^0_{^\cdot}}\right)$. Although through analogies with the $x=0$ case one could directly jump to the result, Eq.~(\ref{F(x)}) was obtained by going through the whole discrete-time analysis. Obviously
\be
L^d F\left[n,\{n_l\}_1^p\right](0)>\int_\Lambda F\left[n,\{n_l\}_1^p\right](x)\d x
\ee
because
\be\label{sum-vs-first_term}
L^d f_n\left(0;\overline{Z^0_{^\cdot}}\right)
=\sum_{z\in\Zz^d}\exp\left\{-\frac{\pi n\lambda_\beta^2}{L^2}(z+\overline{Z^0_{^\cdot}})^2\right\}
>\exp\left\{-\frac{\pi n\lambda_\beta^2}{L^2}\overline{Z^0_{^\cdot}}^2\right\}=\int_\Lambda f_n\left(x;\overline{Z^0_{^\cdot}}\right)\d x.
\ee
If $L\to\infty$, we can distinguish three different domains of $n$. If $n\lambda_\beta^2/L^2\to 0$, the left side of the inequality (\ref{sum-vs-first_term}) diverges while the right side is bounded by 1. If $n\lambda_\beta^2/L^2=c>0$, the left side is $O(1)$ and the right side can remain nonzero. And, when $n\lambda_\beta^2/L^2\to \infty$, the two sides can be asymptotically equal. These outcomes are conditional, they depend on what happens with $\overline{Z^0_{^\cdot}}$ for "typical" values of $\{\alpha^k_j, z^k_j, t^k_{j,r}\}$ in the thermodynamic limit. The typicality will be analyzed in the first part of the proof, and its result will serve to prove the second part of the theorem.

If $n=N$ then $C_l$ ($l\geq 1$) is empty, so $Z^0_1\equiv 0$; if $n<N$, $\delta_{Z^0_1,0}$ could still be dropped from $F\left[n,\{n_l\}_1^p\right](x)$ because $Z^0_1=0$ follows already from $Z^l_1=0$, $l=1,\dots,p$ and $\sum_{l=0}^p Z^l_1\equiv 0$.
From (\ref{F(x)bis}) it is seen that rearrangements of $n_1,\dots,n_p$ do not change the value of $F\left[n,\{n_l\}_1^p\right](x)$.

There is an all-important connection between the Bose-Einstein condensate and the one-particle reduced density matrix
\be\label{sigmaNL1}
\sigma^{N,L}_1=\frac{N}{Q_{N,L}}\Tr_{2,\dots,N}P_+ e^{-\beta H_{N,L}},
\ee
found by Oliver Penrose and Onsager [PO]: the expected number of zero-momentum particles, $\langle N_{0}\rangle$, is equal to the largest eigenvalue of $\sigma^{N,L}_1$.
In (\ref{sigmaNL1}) $P_+$ is the orthogonal projection to the symmetric subspace of the $N$-particle Hilbert space, $H_{N,L}$ is the $N$-particle Hamiltonian on the torus of side $L$ and $\Tr_{2,\dots,N}$ is the partial trace over all but the first particles.
For periodic boundary conditions the spectral resolution of $\sigma^{N,L}_1$ is
\be
\sigma^{N,L}_1=\sum_{k\in (2\pi/L)\Zz^d}\langle N_{k}\rangle |k\rangle\langle k|
\ee
where $|k\rangle\langle k|$ projects to the one-particle state $\langle x|k\rangle=L^{-d/2}e^{\i k\cdot x}$, and $\langle N_{k}\rangle$ is the expected number of particles in this state [S6]. Thus,
\be
\langle N_{0}\rangle=\int_\Lambda \langle x|\sigma^{N,L}_1|0\rangle \d x
\ee
where $\langle x|\sigma^{N,L}_1|0\rangle$ is the integral kernel of $\sigma^{N,L}_1$, and the density of the condensate in finite volumes is
\be\label{rho0/rho}
\rho^{N,L}_0
=\frac{1}{L^d}\int_{\Lambda}\langle x|\sigma^{N,L}_1|0\rangle\d x.
\ee
With the help of the Feynman-Kac formula the integral kernel can be written as
\bea\label{sigma1(0,x1)}
\langle x_1|\sigma^{N,L}_1|0\rangle=\frac{N}{Q_{N,L}N!}\sum_{\pi\in S_N}\int_{\Lambda^{N-1}}\d x_2\cdots\d x_N\int W^\beta_{0x_{\pi(1)}}(\d\omega_1) \int W^\beta_{x_2x_{\pi(2)}}(\d\omega_2)\cdots
\nonumber\\
\cdots\int W^\beta_{x_Nx_{\pi(N)}}(\d\omega_N)\prod_{1\leq j<k\leq N}\exp\left\{-\int_0^\beta u_L(\omega_k(t)-\omega_j(t))\right\}\d t.
\nonumber\\
\eea
Above $S_N$ is the group of permutations of $N$ elements.
The permutations fall into cycles and to each cycle there belongs an effective single-particle trajectory composed of as many physical particles as the length of the cycle. The cycle containing 1 corresponds to an open trajectory that starts in 0 at time 0 and ends in $x_1$ at time $\beta$; all the other trajectories are closed. We therefore treat the cycle of 1 separately and give it the number 0. If the length of this one is $n$ and $\pi$ has $p$ other cycles of lengths $n_1,\dots,n_p$ then the multiple integral in (\ref{sigma1(0,x1)}) is just $e^{-\beta \hat{u}(0)N(N-1)/(2L^d)}F\left[n,\{n_l\}_1^p\right](x_1)$ prior to integration with respect to all but one variables in each cycle. We substitute
\be
\frac{1}{Q_{N,L}}=\frac{L^d\rho^{N,L}_n}{G^N_n},
\ee
rewrite $\pi\in S_N$ as $\pi' \gamma$ where $\gamma$ is the cycle 0 of length $n$ and $\pi'\in S_{N-n}$, and average with respect to $\pi'$ in $S_{N-n}$ to end up with
\be\label{x-sigma-0}
\langle x|\sigma^{N,L}_1|0\rangle=\sum_{n=1}^N\rho^{N,L}_n\, \frac{F^N_n(x)}{F^N_n(0)}
\ee
where $F^N_N(x)=F[N](x)$ and
\be
F^N_n(x)=\sum_{p=1}^{N-n}\frac{1}{p!}\sum_{n_1,\dots,n_p\geq 1:\sum_1^p n_l=N-n}
\frac{F\left[n,\{n_l\}_1^p\right](x)}{\prod_1^p n_l}\quad{\rm if}\quad n=1,\dots,N-1.
\ee
Furthermore,
\be
\rho^{N,L}_0 =\sum_{n=1}^N\rho^{N,L}_n\,
\frac{\int_\Lambda F^N_n(x)\,\d x}
{L^d F^N_n(0)}\ .
\ee
We recall from [S3] the condition that the partition of $N$ sets on $\{\alpha^k_j\}$ via $\prod\delta_{Z^l_1,0}$. Consider a graph ${\cal G}_{\{\alpha^k_j\}}$ of $p+1$ vertices and $\sum_{j\in C_{l'}}\sum_{k\in C_l}\alpha^k_j$ edges between the vertices $0\leq l'<l\leq p$. Then every maximal connected component of ${\cal G}_{\{\alpha^k_j\}}$ which is not an isolated vertex must be a merger through vertices and/or edges of circles of any ($\geq 2$) length. If this does not hold true, $Z^l_1=0$ ($l=0,\dots,p$) cannot be satisfied with nonzero vectors and the corresponding terms are discarded by $\prod\delta_{Z^l_1,0}$. Together with $n$ one can choose first $\alpha^k_j$ for $1\leq j<k\leq n$ and for $1\leq j\leq n< k\leq N$, and then sum over the partitions of $N-n$. Now the partitions are constrained: among them there will remain only those that allow ${\cal G}_{\{\alpha^k_j\}}$ to be a merger of circular graphs. Similarly, choosing first $z^k_{j,r}\neq 0$ for $1\leq j\leq n< k\leq N$ so that $Z^0_1=0$ sets a condition on the other $z^k_{j,r}$ occurring in $Z^l_1$, $l=1,\dots p$ via the identity $\sum_{l=1}^p Z^l_1\equiv -Z^0_1$.
In the proof of the theorem we will follow this way.

\newsec{The ideal Bose gas revisited}\label{ideal}

If there is no interaction, $G\left[\{n_l\}_0^p\right]$ becomes
\be
G^0\left[\{n_l\}_0^p\right]=\prod_{l=0}^p q_{n_l}
\ee
where
\be\label{qn}
q_n=\sum_{z\in\Zz^d}\exp\left\{-\frac{\pi n \lambda_\beta^2}{L^2}z^2\right\}=\frac{L^d}{n^{d/2}\lambda_\beta^d}\sum_{z\in\Zz^d}\exp\left\{-\frac{\pi L^2}{n\lambda_\beta^2}z^2\right\}.
\ee
So $q_n$ is the one-particle partition function at inverse temperature $n\beta$, a monotone decreasing function of $n$ bounded below by 1. Thus, the canonical partition function on the $d$-torus of side $L$ is
\be\label{Q0NL}
Q^0_{N}=\frac{1}{N}\sum_{n=1}^N q_n Q^0_{N-n}.
\ee
Together with the initial condition $Q^0_0=1$, Eq.~(\ref{Q0NL}) defines recursively $Q^0_{N}$. From this equation, which could be obtained without referring to $G\left[\{n_l\}_0^p\right]$, one can easily reproduce most of the results about cycle percolation (the appearance of infinite permutation cycles) and its connection with BEC, obtained in [S1,~2].

With the help of the single-particle energies $\epsilon_k=\hbar^2k^2/2m$, $k\in(2\pi/L)\Zz^d$, the canonical partition function can still be written as
\be\label{Q0N}
Q^0_{N}=\sum_{\sum_{k\neq 0} n_k\leq N}e^{-\beta\sum n_k\epsilon_k}=\sum_{M=0}^N \widehat{Q}^0_M,
\ee
where
\be
\widehat{Q}^0_M=\sum_{\sum_{k\neq 0} n_k=M}e^{-\beta\sum n_k\epsilon_k},
\ee
and the summations run over sets $\{n_k\}_{k\in(2\pi/L)\Zz^d\setminus\{0\}}$ of nonnegative integers. This shows that $Q^0_{N,L}>Q^0_{N-1,L}$, a crucial property for the proof of cycle percolation [S1], which can also be obtained from (\ref{Q0NL}).

\begin{lemma}
Let $A_{-1}=0$, $A_0=1$,
$a_1,a_2,\ldots$ arbitrary numbers, and for $N\geq 1$ define recursively $A_N$ by
\be\label{A}
A_N=\frac{1}{N}\sum_{n=1}^N a_nA_{N-n}.
\ee
Then
\be\label{deltaA}
A_N-A_{N-1}=\frac{1}{N}\sum_{n=1}^N(a_n-1)(A_{N-n}-A_{N-n-1}).
\ee
\end{lemma}

\vspace{5pt}
\noindent
{\em Proof.} This follows by a simple computation. $\quad\Box$

\vspace{5pt}
Now if $a_n>1$, then $A_N>A_{N-1}$ can be proved by induction; and, because $q_n>1$, this applies to $Q_N^0$. Writing Eq.~(\ref{deltaA}) for
$Q^0_N-Q^0_{N-1}=\widehat{Q}^0_N$, keeping only the $n=1$ term and iterating one obtains
\be
\widehat{Q}^0_N> \frac{1}{N}(q_1-1)\widehat{Q}^0_{N-1}>\cdots> \frac{(q_1-1)^N}{N!}\quad(N\geq 1).
\ee
With $\widehat{Q}^0_0=1$ one then concludes that for fixed $L$,
\be
\lim_{N\to\infty}Q^0_N>e^{q_1-1}.
\ee
The limit is finite and is easy to compute from the middle member of (\ref{Q0N}). Because the restriction $\sum_{k\neq 0} n_k\leq N$  drops, the multiple sum factorizes. Using $\beta\epsilon_k=\pi\lambda_\beta^2 z^2/L^2$,
\be
\lim_{N\to\infty}Q^0_N=\prod_{z\in\Zz^d\setminus\{0\}}\left[1-e^{-\pi(\lambda_\beta/L)^2z^2}\right]^{-1} =\exp\left\{\sum_{n=1}^\infty\frac{q_n-1}{n}\right\}\asymp e^{\zeta(1+d/2)(L/\lambda_\beta)^d}\quad (L/\lambda_\beta\gg 1).
\ee
Here $\zeta(x)=\sum_{n=1}^\infty n^{-x}$, the Riemann zeta function. The multiplier of $L^d$ in the exponent is $-\beta$ times the infinite-volume free energy density for the density $\rho$ above its critical value,
\be\label{f0-limit}
f^0(\rho,\beta)=-\frac{\zeta(1+d/2)}{\beta\lambda_\beta^d},\quad \rho\geq \rho^0_c(\beta)=\frac{\zeta(d/2)}{\lambda_\beta^d}.
\ee
Because Eq.~(\ref{deltaA}) has the same form as Eq.~(\ref{A}), some information about $Q^0_{N+1}-2Q^0_N+Q^0_{N-1}$ can also be obtained, e.g., to locate the point where $Q^0_N$ turns from convex to concave, and to show that $L^d\rho^0_c(\beta)$ is on the concave part. (Recall that $\ln Q^0_N$ is concave [LZP], [S4].)

The density of particles in $n$-cycles is now
\be\label{fraction-0}
\rho^{N,L}_n=\frac{q_n Q^0_{N-n}}{L^d Q^0_N}.
\ee

\begin{proposition}\label{thm-ideal-1}
In the ideal Bose gas there is BEC if and only if there exists a $c>0$ such that
\be\label{cond-P0-limit}
\lim_{N,L\to\infty, N/L^d=\rho} \sum_{n>cN^{2/d}}\rho^{N,L}_n>0.
\ee
\end{proposition}

\vspace{5pt}
\noindent
{\em Proof.} (i) Suppose first that (\ref{cond-P0-limit}) holds true. In the absence of interactions
\be
\int_\Lambda f_n(x;0)\d x=1,\quad L^d f_n(0;0)=q_n,
\ee
therefore from Eq.~(\ref{rho0/rho})
\be\label{cond-fraction-ideal}
\rho^{N,L}_0
=\sum_{n=1}^N\frac{\rho^{N,L}_n}{q_n} \geq\sum_{n>cN^{2/d}}\frac{\rho^{N,L}_n}{q_n}
> \frac{\sum_{n>cN^{2/d}}\rho^{N,L}_n}{\sum_{z\in\Zz^d}\exp\{-\pi c(\rho^{1/d}\lambda_\beta)^2 z^2\}},
\ee
where we used the monotonic decrease of $q_n$. Taking the limit we find $\rho_0>0$.

\vspace{10pt}\noindent
(ii) Suppose now that for any $c>0$,
$
\lim_{N,L\to\infty, N/L^d=\rho} \sum_{n>cN^{2/d}}\rho^{N,L}_n=0.
$
We have
\bea
\rho^{N,L}_0=\sum_{n\leq cN^{2/d}}\frac{\rho^{N,L}_n}{q_n} +\sum_{n>cN^{2/d}}\frac{\rho^{N,L}_n}{q_n}
\leq \frac{\sum_{n\leq cN^{2/d}}\rho^{N,L}_n}{\sum_{z\in\Zz^d}\exp\{-\pi c(\rho^{1/d}\lambda_\beta)^2 z^2\}}
+\sum_{n>cN^{2/d}}\rho^{N,L}_n
\nonumber\\
\leq \frac{\rho}{\sum_{z\in\Zz^d}\exp\{-\pi c(\rho^{1/d}\lambda_\beta)^2 z^2\}}
+\sum_{n>cN^{2/d}}\rho^{N,L}_n
\eea
because $q_n>1$. Taking the limit,
\be
\rho_0\leq \frac{\rho}{\sum_{z\in\Zz^d}\exp\{-\pi c(\rho^{1/d}\lambda_\beta)^2 z^2\}}
\ee
for $c$ arbitrarily small, therefore $\rho_0=0$. $\quad\Box$

\vspace{10pt}
This proposition sets a lower bound on the rate of increase of the diverging cycle lengths necessary for the phase transition. Now we prove the phase transition together with the existence of much longer cycles.

\begin{proposition}\label{thm-ideal-2}
For $d\geq 3$ let $\rho\lambda_\beta^d>\zeta(d/2)$, and choose any positive $\varepsilon<1-\frac{\zeta(d/2)}{\rho\lambda_\beta^d}$. Then
\be\label{P-eps}
\lim_{N,L\to\infty,N/L^d=\rho}\sum_{n\geq\varepsilon N}\frac{\rho^{N,L}_n}{\rho}\geq 1-\frac{\zeta(d/2)}{\rho\lambda_\beta^d}-\varepsilon>0.
\ee
\end{proposition}

\vspace{5pt}
\noindent
{\em Remark.}
The point is that we prove cycle percolation, not BEC. That it implies BEC follows from Proposition~\ref{thm-ideal-1}. With the additional information that for $\rho\geq \rho^0_c(\beta)$
\[
\rho-\rho_0=\rho^0_c(\beta)=\frac{\zeta(d/2)}{\lambda_\beta^d}
\]
is the maximum density of particles of nonzero momentum,
\[
1-\frac{\zeta(d/2)}{\rho\lambda_\beta^d}=\frac{\rho_0}{\rho},
\]
so the positive lower bound in (\ref{P-eps}) is the condensate fraction minus $\varepsilon$. Earlier we proved with a more detailed argument that $\geq$ holds with equality,
\be\label{earlier(41)}
\lim_{N,L\to\infty,N/L^d=\rho}\sum_{n>\varepsilon N}\frac{\rho^{N,L}_n}{\rho}=\frac{\rho_0}{\rho}-\varepsilon;
\ee
see [S2, Eq.~(41)].

\vspace{5pt}
\noindent
{\em Proof.} From Eq.~(\ref{fraction-0})
\be
\sum_{n<\varepsilon N}\frac{\rho^{N,L}_n}{\rho}< \sum_{n<\varepsilon N}\frac{q_n}{N}\leq\varepsilon+\frac{1}{\rho}\sum_{n<\varepsilon N}\frac{1}{L^d}\sum_{z\in\Zz^d\setminus\{0\}}e^{-\pi n\lambda_\beta^2 z^2/L^2},
\ee
therefore
\be
\sum_{n\geq\varepsilon N}\frac{\rho^{N,L}_n}{\rho}\geq 1-\varepsilon-\frac{1}{\rho}\sum_{n<\varepsilon N}\frac{1}{L^d}\sum_{z\in\Zz^d\setminus\{0\}}e^{-\pi n\lambda_\beta^2 z^2/L^2},
\ee
whose limit is
\be
\lim_{N,L\to\infty,N/L^d=\rho}\sum_{n\geq\varepsilon N}\frac{\rho^{N,L}_n}{\rho}\geq 1-\varepsilon-\frac{1}{\rho} \sum_{n=1}^\infty\int_{\Rr^d} e^{-\pi n\lambda_\beta^2x^2}\d x=1-\frac{\zeta(d/2)}{\rho\lambda_\beta^d}-\varepsilon. \quad\Box
\ee

\vspace{10pt}
Next, we prove that cycles whose length although diverges but it diverges slower than $N$ have a zero asymptotic density.

\begin{proposition}\label{suppl-ideal}
In any dimension $d\geq 1$ if $K_N\to\infty$ and $K_N/N\to 0$, then
\be\label{slow-divergence}
\lim_{N,L\to\infty,N/L^d=\rho}\sum_{n=1}^{ K_N}\rho^{N,L}_n=\sum_{n=1}^\infty\lim_{N,L\to\infty,N/L^d=\rho}\rho^{N,L}_n\equiv \sum_{n=1}^\infty\rho_n.
\ee
\end{proposition}

\vspace{10pt}
\noindent
{\em Proof.}
We use the rightmost form of $q_n$ in (\ref{qn}).
\be\label{P0NL(KN)}
\sum_{n=1}^{ K_N}\rho^{N,L}_n=\frac{1}{\lambda_\beta^d}\sum_{n=1}^{K_N}\frac{1}{n^{d/2}}\frac{Q^0_{N-n}}{Q^0_N} +\frac{1}{\lambda_\beta^d}\sum_{n=1}^{K_N}\frac{1}{n^{d/2}}\frac{Q^0_{N-n}}{Q^0_N} \sum_{z\neq 0}e^{-\pi L^2z^2/n\lambda_\beta^2}.
\ee
We show that the second sum goes to zero for $K_N=o(N)$ as $L\propto N^{1/d}\to\infty$.
\bea
\sum_{n=1}^{K_N}\frac{1}{n^{d/2}}\frac{Q^0_{N-n}}{Q^0_N} \sum_{z\neq 0}e^{-\pi L^2z^2/n\lambda_\beta^2} &\leq& d\,\frac{\lambda_\beta}{L}\sum_{n=1}^{K_N}\frac{1}{n^{(d-1)/2}} \left[1+2\frac{\sqrt{n}\lambda_\beta}{L}\right]^{d-1}\nonumber\\
&=& \frac{d}{2}\sum_{k=0}^{d-1}{d-1\choose k}\left(\frac{2\,\rho^{1/d}\lambda_\beta}{N^{1/d}}\right)^{d-k}\sum_{n=1}^{K_N}\frac{1}{n^{k/2}}.
\eea
The sum with respect to $n$ is $K_N$ if $k=0$, $O(\sqrt{K_N})$ if $k=1$, $O(\ln K_N)$ if $k=2$, and $O(1)$ if $k\geq 3$.
So for any fixed $M$,
\bea
\sum_{n=1}^M \rho_n
&\leq& \lim_{N,L\to\infty,N/L^d=\rho}\sum_{n=1}^{K_N}\rho^{N,L}_n=\frac{1}{\lambda_\beta^d}\lim_{N,L\to\infty,N/L^d=\rho} \sum_{n=1}^{K_N}\frac{1}{n^{d/2}}\frac{Q^0_{N-n}}{Q^0_N}\nonumber\\
&\leq&\min\left\{\rho,\frac{\zeta(d/2)}{\lambda_\beta^d}\right\}= \min\left\{\rho,\rho^0_c(\beta)\right\}=\sum_{n=1}^\infty\rho_n.
\eea
Taking the limit $M\to\infty$ we obtain Eq.~(\ref{slow-divergence}). $\qquad\Box$

\vspace{10pt}
\noindent
Recall that $\sum_{n=1}^\infty\rho_n$ is the density of particles in finite cycles.
In one and two dimensions $\rho^0_c(\beta)=\infty$, hence $\sum_{n=1}^\infty\rho_n=\rho$. For $d\geq 3$ this proposition tells us that by letting $K_N$ increase slower than $N$ we do not pick up any density coming from infinite cycles.

\vspace{10pt}\noindent
We summarize:

\begin{corollary}\label{thm-ideal-3}
\be\label{ideal-concise}
\rho-\sum_{n=1}^\infty \rho_n
 =\lim_{\varepsilon\downarrow0}\lim_{N,L\to\infty,N/L^d=\rho}\sum_{n>\varepsilon N}\rho^{N,L}_n=\rho_0.
\ee
\end{corollary}
This is the most concise formulation of the connection between cycle percolation and BEC in the ideal Bose gas: the density of particles in infinite cycles is equal to their density in macroscopic cycles, and this is just the condensate density. It then follows that macroscopic cycles do not contribute to the free energy density.
Looking at the expression (\ref{qn}) this seems natural. If there existed trajectories of a diverging length $n=O(N^{2/d})$, they would show up in $f^0$.

Finally, we recall some more results obtained in [S2]. First, the expected number of infinite cycles that contain at least a fraction $x$ of the total number of particles is
$\ln \frac{\rho_0}{x\rho}$ for any $ x<\rho_0/\rho$, cf. Eq.~(44) of [S2]. This number can be arbitrarily large if $x$ is sufficiently small. For $m \geq \ln \rho/\rho_0$ the expected number of infinite cycles of density between
$e^{-(m+1)}\rho$ and $e^{-m}\rho$ is
\[\ln\frac{e^{m+1}\rho_0}{\rho}-\ln\frac{e^{m}\rho_0}{\rho} = 1.\]
The intervals $[e^{-(m+1)}, e^{-m})$ are disjoint, their number is infinite and on average there belongs one infinite cycle to each interval.

Second, the limit shape of partitions of $N$ in Vershik's sense [V] can be inferred from Eqs.~(26), (27) and (44) of [S2]. Let $r_k(\lambda)$ denote the number of elements of length $k$ in the partition $\lambda$ of $N$. Then the limit measure over the set of partitions is singular,
\be
\lim_{N,L\to\infty, N/L^d=\rho}\frac{r_k(\lambda)}{N}= \frac{1}{k}\frac{\rho_k}{\rho}
=\left\{
\begin{array}{lll}
\frac{z^k}{k^{d/2+1}\rho\lambda_\beta^d} & \mbox{if}& \rho\lambda_\beta^d\leq \zeta(d/2)\\
\frac{1}{k^{d/2+1}\rho^0_c(\beta)\lambda_\beta^d}=\frac{1}{k^{d/2+1}\zeta(d/2)} & \mbox{if} & \rho\lambda_\beta^d> \zeta(d/2)
\end{array}
\right.
\ee
with probability one, where $z$ is the solution of
\[
\sum_{n=1}^\infty\frac{z^n}{n^{d/2}}=\rho\lambda_\beta^d.
\]
Thus, with the scaling factor $a=1$ the limit shape for the finite elements of the partitions is
\be
\tilde{\varphi}_\lambda(t):=\frac{a}{N}\sum_{k\geq at}r_k(\lambda)\to\left\{
\begin{array}{lll}
\frac{1}{\rho\lambda_\beta^d}\sum_{k\geq t}\frac{z^k}{k^{d/2+1}} & \mbox{if}& \rho\lambda_\beta^d\leq \zeta(d/2)\\
\frac{1}{\zeta(d/2)}\sum_{k\geq t}\frac{1}{k^{d/2+1}} & \mbox{if} & \rho\lambda_\beta^d> \zeta(d/2).
\end{array}
\right.
\ee
Furthermore, when $\rho\lambda_\beta^d> \zeta(d/2)$, with the scaling sequence $a_N=N\rho^{N,L}_0/\rho$ we obtain the limit shape for the macroscopic elements of the partitions : dropping the prefactor $\rho^{N,L}_0/\rho$,
\be
\tilde{\varphi}^{\rm macr}_\lambda(t)\propto\sum_{k\geq (\rho^{N,L}_0/\rho)Nt}r_k(\lambda)\to \left(\ln\frac{1}{t}\right)_+
\ee
 with probability one.

\newsec{Proof of the Theorem}

Provisionally we strengthen the condition on $\hat{u}$ by assuming that its support is compact, implying also $\int|\hat{u}(x)|x^2\d x<\infty$. Because $u$ is of the positive type, $\hat{u}=|\hat{u}|$.
Let us start by proving a lemma.

\begin{lemma}
For $x$ fixed, when $N, L\to\infty$, $N/L^d=\rho>0$ we have the following asymptotic forms for $f_n\left(x;\overline{Z^0_{^\cdot}}\right)$.

\begin{itemize}
\item[(i)]
 If $n\lambda_\beta^2/L^2\to 0$ then
\be\label{to-0}
f_n\left(x;\overline{Z^0_{^\cdot}}\right)
=
\frac{1}{\lambda_{n\beta}^d}\exp\left\{-\frac{\pi x^2}{n\lambda_\beta^2}\right\}\left[\cos\frac{2\pi}{L}\overline{Z^0_{^\cdot}}\cdot x
+o(1)\right].
\ee
\item[(ii)]
If $n\lambda_\beta^2/L^2\to \infty$ then
\bea\label{to-infty}
f_n\left(x;\overline{Z^0_{^\cdot}}\right)
&=&
\frac{1}{L^d}\exp\left\{-\frac{\pi n\lambda_\beta^2}{L^2}\{\overline{Z^0_{^\cdot}}\}^2\right\}
\nonumber\\
&\times&
\left[\left(1+\sum_{z:\, \max|z_i|=1}\exp\left\{-\frac{\pi n\lambda_\beta^2}{L^2}z\cdot (z+2\{\overline{Z^0_{^\cdot}}\})\right\}   \right)\cos\frac{2\pi}{L}\overline{Z^0_{^\cdot}}\cdot x + o(1)   \right]
\eea
where
$\{\overline{Z^0_{^\cdot}}\}$ is the fractional part of $\overline{Z^0_{^\cdot}}$, each component of which is bounded in modulus by 1/2.
\item[(iii)]
If $n\lambda_\beta^2/L^2=c>0$ then
\be\label{=c}
f_n\left(x;\overline{Z^0_{^\cdot}}\right)=\frac{1}{L^d}\sum_{z\in\Zz^d}\exp\left\{-\pi c(z+\{\overline{Z^0_{^\cdot}}\})^2\right\} \left[\cos\frac{2\pi}{L}\overline{Z^0_{^\cdot}}\cdot x + o(1)\right].
\ee
\end{itemize}
\end{lemma}

\vspace{10pt}
\noindent
{\em Proof.} (i) From the second line of Eq.~(\ref{fnx}),
\bea\label{fn-for-lemma}
f_n\left(x;\overline{Z^0_{^\cdot}}\right)
&=&
\frac{1}{\lambda_{n\beta}^d}
\exp\left\{-\frac{\pi x^2}{n\lambda_\beta^2}\right\}
\nonumber\\
&\times& \left[\cos\frac{2\pi}{L}\overline{Z^0_{^\cdot}}\cdot x
+\sum_{z\in\Zz^d\setminus\{0\}}\exp\left\{-\frac{\pi\left[(x+Lz)^2-x^2\right]}{n\lambda_\beta^2}\right\}\cos\frac{2\pi}{L}\overline{Z^0_{^\cdot}} \cdot (x+Lz)\right].
\nonumber\\
\eea
Because for $z\in\Zz^d$
\[
|x+Lz|^2-x^2\geq L^2 z^2\left(1-2|x|/L\right),
\]
the sum over $\Zz^d\setminus\{0\}$ can be bounded above in modulus by
\[
\sum_{z\in\Zz^d\setminus\{0\}}\exp\left\{-\frac{\pi L^2 z^2}{n\lambda_\beta^2}(1-2|x|/L)\right\}
\]
which tends to zero as $L^2/(n\lambda_\beta^2)$ goes to infinity.

\vspace{5pt}\noindent
(ii) We use the first form of $f_n\left(x;\overline{Z^0_{^\cdot}}\right)$, cf.  Eq.~(\ref{fnx}). With $\overline{Z^0_{^\cdot}}=[\overline{Z^0_{^\cdot}}]+\{\overline{Z^0_{^\cdot}}\}$, a decomposition into integer and fractional parts,
\bea\label{largest-n}
f_n\left(x;\overline{Z^0_{^\cdot}}\right)=\frac{1}{L^d}\sum_{z\in\Zz^d}\exp\left\{-\frac{\pi n\lambda_\beta^2}{L^2}(z+\{\overline{Z^0_{^\cdot}}\})^2\right\}
\cos \frac{2\pi}{L}\left(z-[\overline{Z^0_{^\cdot}}]\right)\cdot x      \hspace{5.5cm}
\nonumber\\
=\frac{1}{L^d}\exp\left\{-\frac{\pi n\lambda_\beta^2}{L^2}\{\overline{Z^0_{^\cdot}}\}^2\right\}
\left[\cos\frac{2\pi}{L}[\overline{Z^0_{^\cdot}}]\cdot x+\sum_{z\in\Zz^d\setminus\{0\}}\exp\left\{-\frac{\pi n\lambda_\beta^2}{L^2}
z\cdot (z+2\{\overline{Z^0_{^\cdot}}\})\right\}   \cos\frac{2\pi}{L}\left(z-[\overline{Z^0_{^\cdot}}]\right)\cdot x   \right].
\nonumber\\
\eea
$z\cdot (z+2\{\overline{Z^0_{^\cdot}}\})\geq 0$, and the sum restricted to  $\max_{1\leq i\leq d}|z_i|\geq 2$ is $o(1)$. The sum over $z$ with $\max|z_i|=1$ can give a contribution of order 1 if $|\{\overline{Z^0_{^\cdot}}\}_i|=1/2$ for one or more components of $\{\overline{Z^0_{^\cdot}}\}$. For these $z$
\[
\cos\frac{2\pi}{L}\left(z-[\overline{Z^0_{^\cdot}}]\right)\cdot x
= \cos\frac{2\pi}{L}\left(\overline{Z^0_{^\cdot}}-\{\overline{Z^0_{^\cdot}}\}-z\right)\cdot x
= \cos\frac{2\pi}{L}\overline{Z^0_{^\cdot}}\cdot x + O(L^{-1}).
\]
Note that $|\{\overline{Z^0_{^\cdot}}\}_i|=1/2$ can only be if the times $t^k_{j,r}$ entering $\overline{Z^0_{^\cdot}}$ take value from a zero-measure subset of $[0,1]^{M}$ where $M=\sum_{j=1}^n\sum_{k=n+1}^N\alpha^k_j$; otherwise the sum over  $z$ with $\max|z_i|=1$ is also $o(1)$.

\vspace{5pt}\noindent
(iii) Looking at the first line of (\ref{largest-n}) it is seen that the summand has a summable majorizing function. Therefore the asymptotic approximation can be done under the summation sign: for any $z$ fixed,
$\cos \frac{2\pi}{L}\left(z-[\overline{Z^0_{^\cdot}}]\right)\cdot x=\cos \frac{2\pi}{L}\overline{Z^0_{^\cdot}}\cdot x + O(L^{-1})$.
$   \Box$

\vspace{5pt}
The essential information provided by the above lemma is that for large systems the $x$-dependence of $f_n\left(x;\overline{Z^0_{^\cdot}}\right)$ and, hence, of $F\left[n,\{n_l\}_1^p\right](x)$ is in the factors
$\exp\left\{-\frac{\pi x^2}{n\lambda_\beta^2}\right\}$ and $\cos\frac{2\pi}{L}\overline{Z^0_{^\cdot}}\cdot x$. From Eqs.~(\ref{to-0})-(\ref{=c})
\be\label{f_n(0)-asymp}
f_n\left(0;\overline{Z^0_{^\cdot}}\right)=
\left\{\begin{array}{lll}
\lambda_{n\beta}^{-d}\,[1+o(1)]&\mbox{if}& n\lambda_\beta^2/L^2\to 0\\
L^{-d}\,e^{-\pi n\lambda_\beta^2\{\overline{Z^0_{^\cdot}}\}^2/L^2}\left[1+\sum_{z:\max|z_i|=1}e^{-\pi n\lambda_\beta^2 z\cdot(z+2\{\overline{Z^0_{^\cdot}}\})}+o(1)\right]&\mbox{if}&n\lambda_\beta^2/L^2\to\infty\\
L^{-d}\sum_{z\in\Zz^d}e^{-\pi c(z+\{\overline{Z^0_{^\cdot}}\})^2}&\mbox{if}&n\lambda_\beta^2/L^2=c.
\end{array}\right.
\ee
Comparison with (\ref{to-0})-(\ref{=c}) shows that
\[
f_n\left(x;\overline{Z^0_{^\cdot}}\right)=
\left\{\begin{array}{lll}
f_n\left(0;\overline{Z^0_{^\cdot}}\right)\exp\left\{-\frac{\pi x^2}{n\lambda_\beta^2}\right\}\left[\cos\frac{2\pi}{L}\overline{Z^0_{^\cdot}}\cdot x+o(1)\right]
&\mbox{if}& n=O(1) \\
f_n\left(0;\overline{Z^0_{^\cdot}}\right)\left[\cos\frac{2\pi}{L}\overline{Z^0_{^\cdot}}\cdot x+o(1)\right]
&\mbox{if}&n\to\infty.\\
\end{array}\right.
\]
However, because $\exp\{-\pi x^2/(n\lambda_\beta^2)\}\to 1$ as $n\to\infty$,
\be\label{fn-unified}
f_n\left(x;\overline{Z^0_{^\cdot}}\right)
=
f_n\left(0;\overline{Z^0_{^\cdot}}\right)
\exp\left\{-\frac{\pi x^2}{n\lambda_\beta^2}\right\}\left[\cos\frac{2\pi}{L}\overline{Z^0_{^\cdot}}\cdot x+o(1)\right]
\ee
holds true in all the cases.

With the help of the lemma the asymptotic form of $F\left[n,\{n_l\}_1^p\right](x)$ is ($n_0=n$)
\bea\label{Fn(x)-asymp}
F\left[n,\{n_l\}_1^p\right](x)
=
\sum_{\alpha^2_1,\alpha^3_1,\alpha^3_2,\dots,\alpha^N_{N-1}=0}^\infty
\, \Delta_{\{\alpha^k_j\},\{n_l\}_0^p}\ \left(L^{-d}\right)^{K_{\{\alpha^k_j\}}}
\prod_{1\leq j<k\leq N}
\frac{\left(-\beta\right)^{\alpha^k_j}}{\alpha^k_j !} \prod_{r=1}^{\alpha^k_j}
 \int_0^1\d t^k_{j,r} \int\d x^k_{j,r}\ \hat{u}\left(x^k_{j,r}\right)
\nonumber\\
\left[\delta(X^0_1,\dots,X^p_1) \prod_{l=1}^p
\exp\left\{-\pi n_l \lambda_\beta^2\left[\overline{\left(X^l_{^\cdot}\right)^2}-\overline{X^l_{^\cdot}}^2\right]\right\}
\sum_{z\in\Zz^d}\exp\left\{-\frac{\pi n_l \lambda_\beta^2}{L^2}\left(z+L\overline{X^l_{^\cdot}}\right)^2\right\}
\right.
\nonumber\\
\left.
\times
\exp\left\{-\pi n \lambda_\beta^2\left[\overline{\left(X^0_{^\cdot}\right)^2}-\overline{X^0_{^\cdot}}^2\right]\right\}
f_n\left(x;L\overline{X^0_{^\cdot}}\right)
\phantom{\prod_{s=S}^T}\hspace{-15pt}\right]
\nonumber\\
\eea
where now $f_n\left(x;L\overline{X^0_{^\cdot}}\right)$ can be substituted from Eq.~(\ref{fn-unified}).
The asymptotic form for $F\left[n,\{n_l\}_1^p\right](0)$ was introduced in [S3].
In it $z^k_{j,r}/L$ is replaced by the continuous variable $x^k_{j,r}$, and $Z_q(t)/L$, $\overline{Z^l_{^\cdot}}/L$, $\overline{(Z^l_{^\cdot})^2}/L^2$, $Z^l_1/L$ are replaced by $X_q(t)$, $\overline{X^l_{^\cdot}}$, $\overline{(X^l_{^\cdot})^2}$, $X^l_1$, respectively. Furthermore,
$L^{-d}\sum_{z^k_{j,r}\in\Zz^d\setminus\{0\}}\hat{u}(z^k_{j,r}/L)$ is replaced by $\int_{\Rr^d}\hat{u}(x^k_{j,r})\d x^k_{j,r}$.
$K_{\{\alpha^k_j\}}$ is the number of linearly independent constraints $X^l_j=0$ and the factor $\left(L^{-d}\right)^{K_{\{\alpha^k_j\}}}$ is implicit in Eq.~(\ref{F(x)}). Here it appears jointly with $\delta(X^0_1,\dots,X^p_1)$ which restricts the integrations with respect to $\{x^k_{j,r}\}$ to a $d\left(\sum_{j<k}\alpha^k_j-K_{\{\alpha^k_j\}}\right)$-dimensional manifold on which every $X^l_1$ is zero. Moreover, $\Delta_{\{\alpha^k_j\},\{n_l\}_0^p}=1$ if ${\cal G}_{\{\alpha^k_j\}}$ is a merger graph of circles, cf. Section 2, and is zero otherwise; now we include this constraint explicitly.

To write down $F^N_n(x)/F^N_n(0)$ we separate the contribution of the $p$ closed trajectories,
\bea\label{HN-n}
H_{N-n}
=\sum_{p=1}^{N-n}\frac{1}{p!}\sum_{n_1,\dots,n_p\geq 1:\sum_1^p n_l=N-n}\frac{1}{\prod_1^p n_l}  \hspace{9cm}
\nonumber\\
\sum_{\{\alpha^k_j\in\Nn_0|n+1\leq j< k\leq N\}} \left(\prod_{n+1\leq j<k\leq N}\frac{\left(-\beta\right)^{\alpha^k_j}}{\alpha^k_j !}\right)
\int_0^1 \prod_{n+1\leq j<k\leq N}\prod_{r=1}^{\alpha^k_j}\d t^k_{j,r}
\int \prod_{n+1\leq j<k\leq N} \prod_{r=1}^{\alpha^k_j}\d x^k_{j,r}\hat{u}\left(x^k_{j,r}\right)
\nonumber\\
\left[\delta(X^0_1,\dots,X^p_1)\,\Delta_{\{\alpha^k_j\},\{n_l\}_0^p}\, \left(L^{-d}\right)^{K_{\{\alpha^k_j\}}}
\prod_{l=1}^p
e^{-\pi n_l \lambda_\beta^2\left[\overline{\left(X^l_{^\cdot}\right)^2}-\overline{X^l_{^\cdot}}^2\right]}\sum_{z\in\Zz^d}e^{-(\pi n_l \lambda_\beta^2/L^2)\left(z+L\overline{X^l_{^\cdot}}\right)^2}\right].
\nonumber\\
\eea
$H_{N-n}$ depends on all the variables that connect the cycles $l=1,\dots,p$ to cycle 0, i.e. on
\[
\{\alpha^k_j,x^k_{j,r},t^k_{j,r}| j\leq n, k\geq n+1, r\leq \alpha^k_j\}.
\]
Let
\bea\label{Theta}
\lefteqn{
\Theta^N_n(y)
=
\sum_{\{\alpha^k_j\in\Nn_0|1\leq j\leq n,\, j+1\leq k\leq N\}} \left(\prod_{ j=1}^n\prod_{k=j+1}^N\frac{\left(-\beta\right)^{\alpha^k_j}}{\alpha^k_j !}\right)
\int_0^1 \prod_{ j=1}^n\prod_{k=j+1}^N\prod_{r=1}^{\alpha^k_j}\d t^k_{j,r}    }
\nonumber\\
&& \int \prod_{j=1}^n\prod_{k=j+1}^N \prod_{r=1}^{\alpha^k_j}\d x^k_{j,r}\, \delta(y-\overline{X^0_{^\cdot}})\hat{u}\left(x^k_{j,r}\right) H_{N-n}
\exp\left\{-\pi n \lambda_\beta^2\left[\overline{\left(X^0_{^\cdot}\right)^2}-y^2\right]\right\}L^d f_n\left(0;Ly\right),
\eea
i.e. the integrations over $\{t^k_{j,r},\,x^k_{j,r}|j\leq n, k\geq j+1, r\leq \alpha^k_j\}$ are restricted to values that yield $\overline{X^0_{^\cdot}}=y$.
Then
\be\label{L^d F^N_n(x)}
L^d F^N_n(x)
=
\exp\left\{-\frac{\pi x^2}{n\lambda_\beta^2}\right\}
\int_{\Rr^d}\Theta^N_n(y)[\cos (2\pi y\cdot x)+o(1)] \d y.
\ee
Introducing
\be\label{nu}
\nu^{N,L}_n(y)=\frac{\Theta^N_n(y)}{L^d F^N_n(0)}=\frac{\Theta^N_n(y)}{\int\Theta^N_n(y')\d y'},
\ee
the ratio $F^N_n(x)/F^N_n(0)$ takes the form
\be\label{Fnx/Fn0}
\frac{F^N_n(x)}{F^N_n(0)}
=
\exp\left\{-\frac{\pi x^2}{n\lambda_\beta^2}\right\}\int_{\Rr^d}[\cos(2\pi y\cdot x)+o(1)]\,\nu^{N,L}_n(y)\d y \,.
\ee
$\int\nu^{N,L}_n(y)\d y=1$, but $\nu^{N,L}_n(y)$ can be negative for some values of $y$. Note also that $\nu^{N,L}_n(y)=\nu^{N,L}_n(-y)$ and, if $u$ is spherically symmetric then $\nu^{N,L}_n$ has a cubic symmetry that tends to spherical as $L$ inreases.
We anticipate the result that $\nu^{N,L}_n(y)$ is an approximation of $\delta(y)$, converging weakly to it as $N, L$ and $n$ with them tend to infinity. The rate of this convergence turns out to be crucial for the proof of the second part of the theorem.

\subsection{ Condition for ODLRO}\label{cond-ODLRO}

The general concept of off-diagonal long-range order was introduced by C. N. Yang [Y]. In the present context it means that
$\sigma_1(x)$ does not go to zero as $x$ goes to infinity.
The condition for ODLRO is slightly weaker than that for BEC; in principle, ODLRO can exist without BEC, but BEC implies ODLRO. In the ideal Bose gas the two notions coincide, in the interacting system this needs a proof.

If in the infinite system the full density arises from particles in finite cycles, i.e. $\rho=\sum_{n=1}^\infty \rho_n$, then we can make use of the uniform upper bound
\be
F\left[n,\{n_l\}_1^p\right](x)\leq \exp\left\{-\frac{\pi x^2}{n\lambda_\beta^2}\right\}  F\left[n,\{n_l\}_1^p\right](0)
\ee
leading to
\be\label{sigma1-from-finite n}
\langle x|\sigma^{N,L}_1|0\rangle \leq \sum_{n=1}^N \rho^{N,L}_n \exp\left\{-\frac{\pi x^2}{n\lambda_\beta^2}\right\}\rightarrow
\sum_{n=1}^\infty \rho_n \exp\left\{-\frac{\pi x^2}{n\lambda_\beta^2}\right\}.
\ee
The infinite sum decays to zero as $x$ goes to infinity. Thus, there is no ODLRO and no BEC either:
\be
\rho^{N,L}_0\leq \sum_{n=1}^N \rho^{N,L}_n \frac{1}{L^d}\int_\Lambda\exp\left\{-\frac{\pi x^2}{n\lambda_\beta^2}\right\}\d x \rightarrow 0.
\ee
In the absence of interaction the inequalities hold with equality.

On the other hand, if
\be\label{cycle-percolation}
\rho-\sum_{n=1}^\infty \rho_n
=
\lim_{M\to\infty} \lim_{N,L\to\infty,N/L^d=\rho}\sum_{n=M}^N\rho^{N,L}_n
>0,
\ee
the infinite-volume Gibbs measure assigns a positive weight to infinite cycles.
When $N$, $L$ and $n$ tend to infinity then for $x$ fixed $\exp\left\{-\frac{\pi x^2}{n\lambda_\beta^2}\right\}$ tends to 1, and
\be\label{Phi}
\frac{F^N_n(x)}{F^N_n(0)}\to\Phi(x)=\int_{\Rr^d}\cos (2\pi y\cdot x)\nu(\d y).
\ee
For the existence of ODLRO $\nu$ must contain a positive multiple of $\delta_0$: $\Phi$ being nonnegative, in the absence of an atomic component at 0 there can be no other atomic component either
so $\nu$ defines a continuous measure, and by the Riemann-Lebesgue lemma $\Phi(x)$ goes to zero when $x$ goes to infinity.
(If $\nu$ is spherically symmetric, a Dirac delta at $|y|=R>0$ in three dimensions adds to $\Phi(x)$ a term $\sim \sin(2\pi R|x|)/(R|x|)$ which is no source for ODLRO, even if we disregard its oscillation.)
In the ideal Bose gas
$\nu=\delta_0$, therefore $\Phi(x)\equiv 1$. Indeed, in this case
\be
F^N_n(x)=\left[ \sum_{p=1}^{N-n}\frac{1}{p!}\sum_{\{n_l\}_1^p\vdash N-n}\prod_{l=1}^p\frac{q_{n_l}}{n_l} \right] f_n(x;0)
\ee
where
\[
f_n(x;0)=f_n(0;0)\,[1+o(1)]\quad\mbox{as}\quad n\to\infty.
\]
Therefore
\[
\frac{F^N_n(x)}{F^N_n(0)}= \frac{f_n(x;0)}{f_n(0;0)}= 1+o(1) \to 1\quad\mbox{as}\quad n\to\infty.
\]
In principle, $\nu$ and $\Phi$ may depend on how fast $n$ increases with $N$. It is better to treat $n$ as an independent variable, because $N\mapsto n(N)$ cannot be strictly increasing unless $N-n$ is a constant. Inverting it we obtain $N={\cal N}(n)$ that strictly increases through jumps. The fastest increase of $n$ that we shall consider here is $n\sim\varepsilon N$ with some $0<\varepsilon<1$, so the slowest increase for ${\cal N}$ is ${\cal N}(n)=\lfloor n/\varepsilon\rfloor$. Accordingly, $L$ must satisfy ${\cal N}(n)/L^d=\rho$. It will turn out that what counts is only that $n$ goes to infinity, the way as ${\cal N}(n)$ increases has a subordinate effect on $\nu$ and $\Phi$. While $\beta$ will appear in the asymptotic form of $\nu^{N,L}_n$, no prognostic about its critical value can be inferred from the formula. The crucial role of $\rho$ and $\beta$ is elsewhere, they decide about the fulfilment of the inequality (\ref{cycle-percolation}).

To see whether $\nu$ contains a multiple of $\delta_0$, we must analyse the large-$n$ behavior of $\overline{X^0_{^\cdot}}$ and $\overline{\left(X^0_{^\cdot}\right)^2}$. From Eq.~(\ref{avZ-0})
\be\label{avX_0}
n\overline{X^0_{^\cdot}}= -\sum_{k=2}^n\sum_{j=1}^{k-1}(k-j)\sum_{r=1}^{\alpha^{k}_{j}}x^{k}_{j,r}
+\sum_{j=1}^{n}\sum_{k=1}^{N-n}\sum_{r=1}^{\alpha^{+k}_{j}}\left(j-1+t^{+k}_{j,r}\right)x^{+k}_{j,r}
=: nY^0_0 +nY^+_0.
\ee
In the new notation we restart the count of the particles outside the zeroth cycle. Now cycle 0 is distinguished without making distinction among the different partitions of $N-n$. Similarly, from [S3,~Eqs.~(3.58), (3.59)] with a more detailed notation
\bea\label{avZl-2}
n\overline{\left(X^0_{^\cdot}\right)^2}
&=&\sum_{1\leq j<k\leq n}\ \sum_{1\leq j'<k'\leq n}\sum_{r=1}^{\alpha^k_j}\sum_{r'=1}^{\alpha^{k'}_{j'}}A^{k'j'r'}_{kjr}x^k_{j,r}\cdot x^{k'}_{j',r'}
\nonumber\\
&+&\sum_{j=1}^{n}\sum_{k=1}^{N-n} \sum_{j'=1}^{n}\sum_{k'=1}^{N-n}
\sum_{r=1}^{\alpha^{+k}_j}\sum_{r'=1}^{\alpha^{+k'}_{j'}}A^{+k'j'r'}_{+kjr} x^{+k}_{j,r}\cdot x^{+k'}_{j',r'}
-2\sum_{1\leq j<k\leq n}\sum_{j'=1}^{n}\sum_{k'=1}^{N-n} \sum_{r=1}^{\alpha^k_j}\sum_{r'=1}^{\alpha^{+k'}_{j'}}A^{+k'j'r'}_{kjr} x^k_{j,r}\cdot x^{+k'}_{j',r'}
\nonumber\\
\eea
where
\be
A^{+k'j'r'}_{+kjr}=\min\left\{j-1+t^{+k}_{j,r},\,j'-1+t^{+k'}_{j',r'}\right\},
\ee
\be
A^{+k'j'r'}_{kjr}=\left\{\begin{array}{lll}
k-j&\mbox{if}&k<j'\\
j'-j+t^{+k'}_{j',r'}-t^{k}_{j,r}&\mbox{if}&j< j'<k\\
0&\mbox{if}&j'<j\ ,
\end{array}\right.
\ee
\be
A^{k'j'r'}_{kjr}=\left\{\begin{array}{lll}
0&\mbox{if}&j'<k'< j<k\ \ \mbox{or}\ \ j<k< j'<k'\\
k-j&\mbox{if}&j'< j<k< k'\\
k'-j'&\mbox{if}&j< j'<k'< k\\
k-j'+t^k_{j,r}-t^{k'}_{j',r'}&\mbox{if}&j<j'< k<k'\\
k'-j+t^{k'}_{j',r'}-t^{k}_{j,r}&\mbox{if}&j'< j<k'< k\ .\\
\end{array}\right.
\ee
If two indices coincide, there may appear a correction in modulus not larger than 1, coming from the times $t^k_{j,r}, t^{+k}_{j,r}$, but there is no such correction to $A^{kjr}_{kjr}=k-j$.

Because $\overline{X^0_{^\cdot}}$ and $\overline{\left(X^0_{^\cdot}\right)^2}$ depend on all the variables $\{\alpha^k_j,\,t^k_{j,r},\,x^k_{j,r}|j\leq n, k\geq j+1, r\leq \alpha^k_j\}$, it is advantageous to rearrange the summations in $\Theta^N_n$. Let
\be\label{vector-alpha}
~\alpha^0_n=\{\alpha^k_j|1\leq j< k\leq n\},\quad ~\alpha^+_n=\{\alpha^{+k}_j|1\leq j\leq n,\, 1\leq k\leq {\cal N}(n)-n\},\quad ~\alpha_n=\{~\alpha^0_n,\,~\alpha^+_n\}.
\ee
Let, moreover,
\be\label{alpha-norms}
\|~\alpha^0_n\|=\sum_{k=2}^n\sum_{j=1}^{k-1} j\  \alpha^k_{k-j},\quad \|~\alpha^+_n\|=\sum_{k=1}^{{\cal N}(n)-n}\sum_{j=1}^n \ j\alpha^{+k}_j,
\quad \|~\alpha_n\|=\|~\alpha^0_n\|+\|~\alpha^+_n\|.
\ee
Then
\be
\Theta^N_n(y)=\sum_{A,B=0}^\infty\,\sum_{~\alpha^0_n:\|~\alpha^0_n\|=A}\,\sum_{~\alpha^+_n:\|~\alpha^+_n\|=B}\vartheta_{~\alpha_n}(y)
\ee
where $\vartheta_{~\alpha_n}(y)$ is defined by Eq.~(\ref{Theta}).
For most of what follows we shall consider the infinite sequences $~\alpha^0=\{\alpha^k_j|1\leq j<k<\infty\}$ and $~\alpha^+=\{\alpha^{+k}_j| j, k\in \Nn_+\}$ as given beforehand, and then $~\alpha^0_n$ and $~\alpha^+_n$ are their restrictions to a finite number of elements as shown in Eq.~(\ref{vector-alpha}).
Since $t^k_{j,r}$ and $t^{+k}_{j,r}$ play a minor role, they can also be considered as given beforehand.

In points (i)-(v) below we identify the $~\alpha=\{~\alpha^0, ~\alpha^+\}$ that give rise to a decaying $\overline{X^0_{^\cdot}}$.

\vspace{3pt}
\noindent
(i) If $~\alpha$ contains a finite number of nonzero elements then $\|~\alpha_n\|$ attains a constant at a finite $n$, implying
$\overline{X^0_{^\cdot}}\to 0$. We shall refer to these sequences as rationals,
while those with infinitely many positive elements will be referred to as irrationals.
With all the $x^k_{j,r}$ and $x^{+k}_{j,r}$ on the support of $\hat{u}$ (that we suppose in the sequel) $n\overline{X^0_{^\cdot}}$ and $n\overline{\left(X^0_{^\cdot}\right)^2}$ are bounded, so $|\overline{X^0_{^\cdot}}|=O(1/n)$ and
\[
n\left[\overline{\left(X^0_{^\cdot}\right)^2}-\overline{X^0_{^\cdot}}^2\right]
= n\overline{\left(X^0_{^\cdot}\right)^2}-O(1/n)=O(1).
\]

\noindent
(ii) $\overline{X^0_{^\cdot}}$ tends surely to zero also for some irrational sequences.
If  $\|~\alpha_n\|\to\infty$ but
\be\label{cond-sure-convergence}
\frac{\|~\alpha_n\|}{n}
=\frac{1}{n}\left[ \sum_{k=2}^n\sum_{j=1}^{k-1} j\  \alpha^k_{k-j}
+\sum_{k=1}^{{\cal N}(n)-n}\sum_{j=1}^n \ j\alpha^{+k}_j\right] \to 0
\ee
then $\overline{X^0_{^\cdot}}$ goes to zero, although the decay is slower than $1/n$. An example is $~\alpha^+=~0$, $\alpha^k_{k-1}=1$ if $k$ is a prime and $\alpha^k_j=0$ otherwise. Concerning the asymptotic form of $n\left[\overline{\left(X^0_{^\cdot}\right)^2}-\overline{X^0_{^\cdot}}^2\right]$ more information is necessary and will be obtained later.

\vspace{3pt}
\noindent
(iii) To enlarge further the family of irrational sequences yielding $\overline{X^0_{^\cdot}}\to 0$ and also to get some insight into the asymptotic form of $\overline{\left(X^0_{^\cdot}\right)^2}-\overline{X^0_{^\cdot}}^2$ we use probabilistic arguments.
The vectors $x^k_{j,r}$ and $x^{+k}_{j,r}$ can be considered as identically distributed zero-mean random variables with the common probability density $|\hat{u}(\cdot)|/\|\hat{u}\|_1$. Viewed as such, $n\overline{X^0_{^\cdot}}$ is the result of a finite number of steps of a symmetric random walk in $\Rr^d$.
The vectors $x^k_{j,r}$ are independent among themselves and from  $x^{+k}_{j,r}$, but the latter are weakly dependent because their sum vanishes,
\be\label{x+1,11}
X^0_1=\sum_{j=1}^n\sum_{k=1}^{N-n}\sum_{r=1}^{\alpha^{+k}_j}x^{+k}_{j,r}=0.
\ee
One vector, e.g. $x^{+1}_{1,1}$ can be expressed with the others, yielding
\be\label{Y+0-indept}
n Y^+_0=\sum_{j=1}^{n}\sum_{k=1}^{N-n}\sum_{r=1}^{\alpha^{+k}_{j}}\left(j-1+t^{+k}_{j,r}-t^{+1}_{1,1}\right)x^{+k}_{j,r}.
\ee
The vectors appearing here with a nonzero coefficient are already independent. The product measures for finite $N, L$ form a consistent family and $\Pp$ is their unique extension to a probability measure in the measurable space $\left((\Rr^d)^\infty, \Sigma\right)$, where $\Sigma$ is the smallest $\sigma$-algebra containing all sets that depend only on a finite number of $x^k_{j,r}$, $x^{+k}_{j,r}$. In a short-hand notation $F^N_n(0)$ can be rewritten as
\be
F^{{\cal N}(n)}_n(0)=
\sum_{~\alpha_n} \frac{(-\beta \|\hat{u}\|_1)^{~\alpha_n}}{~\alpha_n!} \int \d t_{~\alpha_n}\,
\Ee\left[H_{{\cal N}(n)-n}\exp\left\{-\pi n \lambda_\beta^2\left[\overline{\left(X^0_{^\cdot}\right)^2}-\overline{X^0_{^\cdot}}^2\right]\right\}
f_n\left(0;L\overline{X^0_{^\cdot}}\right) \right].
\ee
Now $\overline{X^0_{^\cdot}}\to 0$
is a tail event, so it occurs with probability 0 or 1.
In the probabilistic setting the convergence of $Y^0_0$ to zero is an instance of the strong law of large numbers and it holds with probability one e.g. for the sequence
\be\label{intracycle}
\alpha^k_{k-j}=\left\{\begin{array}{ll}
a & {\rm if}\ j\leq j_0, k=2, 3,\dots\\
0 & {\rm otherwise}
\end{array}\right.
\ee
where $a, j_0\geq 1$. This can be seen by expressing $Y^0_0$ as an average of independent zero-mean random variables,
\be\label{lambda-u}
Y^0_0=\frac{1}{n}\sum_{k=2}^n\xi_k\quad\mbox{with}\quad \xi_k=-\sum_{j=1}^{k-1}j\sum_{r=1}^{\alpha^k_{k-j}}x^k_{k-j,r}\quad\mbox{and}\quad
\Ee[\xi_k^2]=\frac{1}{\lambda_u^2} \sum_{j=1}^{k-1}j^2\alpha^k_{k-j}.
\ee
Here the length $\lambda_u$ is defined by $1/\lambda_u^2=\int|\hat{u}(x)|x^2\d x/\|\hat{u}\|_1$.
The condition
\be\label{conditionF}
\sum_{k=1}^\infty \frac{1}{k^2}\Ee[\xi_k^2]
<\infty
\ee
implies that with probability one $\sum_{k=1}^\infty k^{-1}\xi_k$ is convergent and $n^{-1}\sum_{k=1}^n\xi_k\to 0$ [Fe], and is satisfied by (\ref{lambda-u}). So for $~\alpha^0$ defined by Eq.~(\ref{intracycle}) $\|~\alpha^0_n\|\propto n$ and $Y^0_0\to 0$ almost surely. The latter is true also if in (\ref{intracycle}) $k\in\Nn_+$ is replaced by $k$ being in a positive- or zero-density subset of $\Nn_+$. We shall return to this case later. In these examples $\xi_k$ is a bounded sequence (for $x^k_{k-j,r}\in\supp \hat{u}$). The condition (\ref{conditionF}) may hold also if in (\ref{lambda-u}) $\xi_k$ is unbounded, e.g., if
\be\label{alpha-in}
\alpha^k_{k-j}=\left\{\begin{array}{lll}
a & {\rm if} & j\leq  k^\theta, k=2, 3,\dots\\
0 & {\rm if} & j>k^\theta
\end{array}\right.
\ee
provided that $\theta<1/3$. In this case $\|~\alpha^0_n\|/n$ diverges as $n^{2\theta}$, still $Y^0_0\to 0$ almost surely.

To apply the strong law of large numbers to $Y^+_0$, we extend the summation with respect to $k$ up to infinity and introduce
\[
\eta_j=\sum_{k=1}^\infty\sum_{r=1}^{\alpha^{+k}_j} \left(j-1+t^{+k}_{j,r}-t^{+1}_{1,1}\right)x^{+k}_{j,r},\quad V_n=\frac{1}{n}\sum_{j=1}^n\eta_j.
\]
Then
\[
\Ee[\eta_j^2]=\frac{1}{\lambda_u^2}\sum_{k=1}^\infty\sum_{r=1}^{\alpha^{+k}_j} \left(j-1+t^{+k}_{j,r}-t^{+1}_{1,1}\right)^2,
\]
and
\[
\sum_{j=1}^\infty\frac{1}{j^2}\Ee[\eta_j^2]=\frac{1}{\lambda_u^2}\sum_{k=1}^\infty\sum_{r=1}^{\alpha^{+k}_j}
\frac{ \left(j-1+t^{+k}_{j,r}-t^{+1}_{1,1}\right)^2}{j^2}
\leq \frac{1}{\lambda_u^2}\sum_{k=1}^\infty\alpha^{+k}_j<\infty
\]
if $\alpha^{+k}_j>0$ only for a finite number of pairs $(j,k)$, i.e. if $~\alpha^+$ is rational. This guaranties the sure convergence of $V_n$ and also of $Y^+_0$ to zero, but the result is not new.

In Eq.~(\ref{lambda-u}) we wrote $Y^0_0$ as $1/n$ times the resulting vector of $n$ steps of a random walk. If the step lengths are bounded then the averaged distancing from the origin must decay as $1/\sqrt{n}$. The rest of the proof is devoted to show that in the dominant contribution to the partition function the step lengths are indeed bounded.

\vspace{5pt}
\noindent
(iv) More systematically, we can find irrational sequences giving rise to $\overline{X^0_{^\cdot}}\to 0$ a.s. as follows. Note first that if $\|~\alpha_n\|/n\to\infty$ then $\overline{X^0_{^\cdot}}$ can attain any $y\in\Rr^d$ as $n\to\infty$, but $\Pp(\overline{X^0_{^\cdot}}\to y)=0$ if $y\neq 0$: because $\Ee[\overline{X^0_{^\cdot}}]=0$, for $y\neq 0$ and $\kappa<|y|$ fixed,  $|\overline{X^0_{^\cdot}}-y|<\kappa$ becomes an event of large deviation as $n$ goes to infinity. This implies that for the given $~\alpha$ the asymptotic probability distribution is continuous outside 0, but it is not excluded that also $\Pp(\overline{X^0_{^\cdot}}\to 0)=0$. On the other hand, $\Ee\left[\overline{X^0_{^\cdot}}^2\right]\to 0$ already implies $\Pp(\overline{X^0_{^\cdot}}\to 0)=1$, because $\overline{X^0_{^\cdot}}^2>s>0$ is a large deviation from $\Ee\left[\overline{X^0_{^\cdot}}^2\right]$ for $n$ large enough.
However, as shown below, arguing with $\Ee\left[\overline{X^0_{^\cdot}}^2\right]\to 0$ will not significantly extend the result of points (i)-(iii).
\[
\Ee\left[\overline{X^0_{^\cdot}}^2\right]
=\Ee\left[\left(Y^0_0\right)^2\right] + \Ee\left[\left(Y^+_0\right)^2\right]
\]
where
\be\label{EY0-EY+}
\Ee\left[\left|Y^0_0\right|^2\right]=\frac{1}{n^2\lambda_u^2} \sum_{k=2}^n \sum_{j=1}^{k-1}j^2\alpha^k_{k-j},\quad
\Ee\left[\left|Y^+_0\right|^2\right]=\frac{1}{n^2\lambda_u^2}\sum_{j=1}^{n}\sum_{k=1}^{{\cal N}(n)-n}\ \sum_{r=1}^{\alpha^{+k}_{j}}\left(j-1+t^{+k}_{j,r}-t^{+1}_{1,1}\right)^2.
\ee
Hence, $\overline{X^0_{^\cdot}}^2$ goes to zero with probability one if
\be\label{E[.]->0}
\frac{1}{n^2}\sum_{k=2}^n \sum_{j=1}^{k-1} j^2 \alpha^k_{k-j}\to 0
\quad\mbox{and}\quad
\frac{1}{n^2} \sum_{k=1}^{{\cal N}(n)-n}\sum_{j=1}^{n}\ \sum_{r=1}^{\alpha^{+k}_{j}}\left(j-1+t^{+k}_{j,r}-t^{+1}_{1,1}\right)^2\to 0.
\ee
Equation (\ref{E[.]->0}) can hold if $\|~\alpha_n\|/n$ does not go to zero or even tends to infinity. For example, the first condition is satisfied for $~\alpha^0$ defined by (\ref{intracycle}) or (\ref{alpha-in}); thus, one can reproduce the result obtained from the strong law of large numbers. With (\ref{intracycle}), $\|~\alpha_n\|/n\sim 1$ but
\[\Ee\left[\left(Y^0_0\right)^2\right]\propto \frac{1}{n^2}\sum_{k=2}^n \sum_{j=1}^{\min\{k-1, j_0\}} j^2 \alpha^k_{k-j}=O\left(1/n\right);\]
with (\ref{alpha-in}), $\|~\alpha_n\|/n\sim n^{2\theta}$
but
\[\Ee\left[\left(Y^0_0\right)^2\right]\propto \frac{1}{n^2}\sum_{k=2}^n \sum_{j=1}^{\min\{k-1, k^\theta\}} j^2 \alpha^k_{k-j}=O\left(\frac{1}{n^{1-3\theta}}\right)\to 0\]
if $\theta<1/3$.

The second condition in (\ref{E[.]->0}) is also fulfilled by some irrational sequences.
Suppose first that ${\cal N}(n)/n^2\to 0$ as $n\to\infty$, for example, $n=\varepsilon N$. If for any $k$
\[
\alpha^{+k}_{j}=\left\{\begin{array}{lll}
a & {\rm if} & j\leq  j_0\\
0 & {\rm if} & j>j_0.
\end{array}\right.
\]
then
\be\label{alpha+k_j}
\frac{1}{n^2} \sum_{k=1}^{{\cal N}(n)-n}\ \sum_{j=1}^{n}\ \sum_{r=1}^{\alpha^{+k}_{j}}\left(j-1+t^{+k}_{j,r}-t^{+1}_{1,1}\right)^2
\leq \frac{j_0(j_0+1)(2j_0+1)a}{6}\ \frac{{\cal N}(n)-n}{n^2}\to 0.
\ee
If ${\cal N}(n)/ n^2$ does not go to zero, let $\{k_i\}$ be a lacunary sequence such that $n^{-2}\sum_{i:\ k_i\leq {\cal N}(n)-n}1\to 0$. Then for
\[
\alpha^{+k}_{j}=\left\{\begin{array}{ll}
a & {\rm if}\ j\leq  j_0,\ k\in\{k_i\}\\
0 & {\rm otherwise}
\end{array}\right.
\]
the condition is satisfied, $Y^+_0\to 0$ with probability one. This is new but irrelevant.
The statistical weight of $~\alpha$ -- rational or irrational, producing $\overline{X^0_{^\cdot}}\to 0$ or not -- is decreased by a factor $\left(L^{-d}\right)^{K_{~\alpha}}$ in which $K_{~\alpha}$ (depending only on $\alpha^k_j$ where $j$ and $k$ are in different cycles) is
at least as large as the number of cycles which the zeroth cycle is coupled with (cf. [S3, Remark 6]).
This suggests that the asymptotic contribution of each cycle to $K_{~\alpha}$ is finite; or, in graph language, the infinite graph ${\cal G}_{~\alpha}$ is almost surely of finite degree, whether or not there exist infinite cycles. Later we shall see that more is true: $~\alpha^+$ (and its analogues for cycles $1,\dots,p$) must be rational.

\noindent
(v) A similar comparison can be done between $\overline{(X^0_{^\cdot})^2}$ and $\Ee\left[\overline{(X^0_{^\cdot})^2}\right]$. The convergence of the latter to zero implies $\Pp\left(\overline{(X^0_{^\cdot})^2}\to 0\right)=1$, because if $n$ is large enough, $\overline{(X^0_{^\cdot})^2}>s>0$ is a large deviation from $\Ee\left[\overline{(X^0_{^\cdot})^2}\right]$. As in point (iv), this will not provide us with a new case of $\overline{X^0_{^\cdot}}\to 0$.
Substituting $x^{+1}_{1,1}$ from Eq.~(\ref{x+1,11}) into Eq.~(\ref{avZl-2}) and taking the expectation value,
\bea
\Ee\left[\overline{(X^0_{^\cdot})^2}\right]
=
\frac{1}{n\lambda_u^2}\left[ \sum_{k=2}^n\sum_{j=1}^{k-1}j\alpha^k_{k-j}
+ \sum_{k=1}^{{\cal N}(n)-n}\sum_{j=2}^n\ \sum_{r=1}^{\alpha^{+k}_j}\left(j-1+t^{+k}_{j,r}-t^{+1}_{1,1}\right)
\right.
\nonumber\\
\left.
+\sum_{k=1}^{{\cal N}(n)-n}\ \sum_{r=1}^{\alpha^{+k}_1}\left(t^{+k}_{1,r}+t^{+1}_{1,1}-2\min\left\{t^{+k}_{1,r},t^{+1}_{1,1}\right\}\right)\right]
\nonumber\\
=
\frac{1}{n\lambda_u^2}\left[ \sum_{k=2}^n\sum_{j=1}^{k-1}j\alpha^k_{k-j}+ \sum_{k=1}^{{\cal N}(n)-n}\sum_{j=1}^n\ \sum_{r=1}^{\alpha^{+k}_j}\left|j-1+t^{+k}_{j,r}-t^{+1}_{1,1}\right| \right]
\approx
\frac{1}{n\lambda_u^2}\|~\alpha_n\|.
\eea
Thus, $\Ee\left[\overline{(X^0_{^\cdot})^2}\right]\to 0$ is equivalent to (\ref{cond-sure-convergence}) which was the condition for $\overline{X^0_{^\cdot}}\to 0$ surely.

\vspace{3pt}
\noindent
(vi) To summarize, if $~\alpha$ is rational, the result is deterministic, $|\overline{X^0_{^\cdot}}|=O(1/n)$ and $\overline{\left(X^0_{^\cdot}\right)^2}=O(1/n)$. If $~\alpha$ is irrational and $\|~\alpha_n\|/n\to 0$, the result is still deterministic, both $\overline{X^0_{^\cdot}}$ and $\overline{\left(X^0_{^\cdot}\right)^2}$ go surely to zero. We found also other irrational $~\alpha$ with non-decaying or even diverging $\|~\alpha_n\|/n$
that allow for an $\overline{X^0_{^\cdot}}$ almost surely converging to zero. However, as we shall see, while $\|~\alpha_n\|\sim n$ provides the leading-order contribution, no $~\alpha$ with $\|~\alpha_n\|$ diverging faster than $n$ contributes asymptotically to $\nu^{N,L}_n$.

By inspecting
\bea\label{variance}
\lefteqn{
\Ee\left[\overline{\left(X^0_{^\cdot}\right)^2}-\overline{X^0_{^\cdot}}^2\right]
=
\frac{1}{n\lambda_u^2}
\left[\sum_{k=2}^n\sum_{j=1}^{k-1}j\alpha^k_{k-j}\left(1-\frac{j}{n}\right)
\right. }\nonumber\\
&&+
\left.
\sum_{k=1}^{{\cal N}(n)-n}\sum_{j=1}^n\sum_{r=1}^{\alpha^{+k}_j}\left|j-1+t^{+k}_{j,r}-t^{+1}_{1,1}\right|\left(1-\frac{\left|j-1+t^{+k}_{j,r}-t^{+1}_{1,1}\right|}{n}\right)\right]
\eea
one observes that, unless $\alpha^k_{k-j}> 0$ for $k$ and $j$ of order $n$, $\Ee\left[\overline{\left(X^0_{^\cdot}\right)^2}-\overline{X^0_{^\cdot}}^2\right]$ is of the same order as $\Ee\left[\overline{\left(X^0_{^\cdot}\right)^2}\right]$.
The exception is illustrated by the following example. Let $~\alpha^+_n=~0$, $k_0\geq 1$, $\alpha^{k}_{k-j}=1$ for $n-k_0\leq j<k\leq n$ and $\alpha^k_{k-j}=0$ otherwise. Then $\|~\alpha^0_n\|=\sum_{n-k_0\leq j<k\leq n}j$ is of order $n$, so both $\Ee\left[\overline{\left(X^0_{^\cdot}\right)^2}\right]$ and $\Ee\left[\overline{X^0_{^\cdot}}^2\right]$ are of order 1, but their difference computed from the first line of (\ref{variance}) is of order $1/n$. Later we return to this example, here we note only that $~\alpha^0_n$ is exceptional also because it is not the restriction of some infinite $~\alpha^0$, since $~\alpha^0_n\to ~0$ as $n\to\infty$.

Focusing first on the "regular" case $\Ee\left[\overline{\left(X^0_{^\cdot}\right)^2}\right]\leq C\left(\Ee\left[\overline{\left(X^0_{^\cdot}\right)^2}-\overline{X^0_{^\cdot}}^2\right] \right)$ with some $C>1$ we find that
\be\label{typical}
\lim_{n\to\infty}\frac{n}{\|~\alpha_n\|}\Ee\left[\overline{X^0_{^\cdot}}^2\right]
= c
<\frac{1}{\lambda_u^2}=\Ee\left[(x^k_{j,r})^2\right]=\lim_{n\to\infty}\frac{n}{\|~\alpha_n\|}\Ee\left[\overline{\left(X^0_{^\cdot}\right)^2}\right],
\ee
that is, the random variables $\frac{n}{\|~\alpha_n\|}\left[\overline{\left(X^0_{^\cdot}\right)^2}-\overline{X^0_{^\cdot}}^2\right]$ have a positive bounded expectation value tending to $\lambda_u^{-2}-c>0$ as $n$ goes to infinity.

To see the dominant contribution to $\Theta^{{\cal N}(n)}_n(y)$ one must estimate the weight of all the $~\alpha^0_n$ and $~\alpha^+_n$ with a given norm. This weight is composed of their number multiplied by $\prod (\beta\|\hat{u}\|_1)^{\alpha^k_j}/\alpha^k_j!$, and can compensate the loss due to a decaying
$\exp\left\{-\pi n \lambda_\beta^2\left[\overline{\left(X^0_{^\cdot}\right)^2}-\overline{X^0_{^\cdot}}^2\right]\right\}$.
Consider first $~\alpha^0_n$. As a typical example, let $\alpha^k_{k-j}=a\geq 1$ for $j=1,\dots,j_{~\alpha^0_n}$ and for $i_{~\alpha^0_n}$ different values of $k$ (each larger than $j_{~\alpha^0_n}$), and $\alpha^k_{k-j}=0$ otherwise. Then
\[
\|~\alpha^0_n\|=\frac{a}{2} i_{~\alpha^0_n}j_{~\alpha^0_n}(j_{~\alpha^0_n}+1)
\]
and the associated weight is
\[
\left[\frac{(\beta\|\hat{u}\|_1)^a}{a!}\right]^{i_{~\alpha^0_n}j_{~\alpha^0_n}}{n-j_{~\alpha^0_n}\choose i_{~\alpha^0_n}}
\]
whose logarithm
\bea\label{weight}
\ln \left\{ \left[\frac{(\beta\|\hat{u}\|_1)^a}{a!}\right]^{i_{~\alpha^0_n}j_{~\alpha^0_n}}{n-j_{~\alpha^0_n}\choose i_{~\alpha^0_n}} \right\}
&\approx&
a\, i_{~\alpha^0_n}j_{~\alpha^0_n}(\ln \beta\|\hat{u}\|_1-\ln a+1)-\frac{1}{2}i_{~\alpha^0_n}j_{~\alpha^0_n}\ln 2\pi a+\ln {n-j_{~\alpha^0_n}\choose i_{~\alpha^0_n}}
\nonumber\\
&=&
\|~\alpha^0_n\| \frac{2}{j_{~\alpha^0_n}+1}\left(\ln\frac{\beta\|\hat{u}\|_1}{a}+1-\frac{\ln 2\pi a}{2a}\right)
+\ln {n-j_{~\alpha^0_n}\choose i_{~\alpha^0_n}}
\eea
is to be added to $-\pi \lambda_\beta^2 n \left[\overline{\left(X^0_{^\cdot}\right)^2}-\overline{X^0_{^\cdot}}^2\right]$.

\vspace{3pt}
\noindent
(vii) We start by proving that if $\Ee\left[\overline{\left(X^0_{^\cdot}\right)^2}-\overline{X^0_{^\cdot}}^2\right]\propto \|~\alpha_n\|/n$
goes to infinity, the contribution to the partition function is negligible.
Assume that $\|~\alpha_n\|\sim \|~\alpha^0_n\|$; we shall see that this is indeed the case, the contribution of $\|~\alpha^+_n\|$ to $\|~\alpha_n\|$ is finite. For a reference we note that adding
$
-\pi n\lambda_\beta^2 \Ee\left[\overline{\left(X^0_{^\cdot}\right)^2}-\overline{X^0_{^\cdot}}^2\right]
$
to (\ref{weight})
yields the averaged exponent
\be\label{E-av}
E_{\rm av}=-\|~\alpha^0_n\|\left[\pi\lambda_\beta^2 \left(\lambda_u^{-2}-c\right)
-\frac{2}{j_{~\alpha^0_n}+1}\left(\ln\frac{\beta\|\hat{u}\|_1}{a}+1-\frac{\ln 2\pi a}{2a}\right)\right]
+\ln {n-j_{~\alpha^0_n}\choose i_{~\alpha^0_n}}.
\ee
Because $i_{~\alpha^0_n}\leq n$, $\|~\alpha^0_n\|/n=a\, i_{~\alpha^0_n}j_{~\alpha^0_n}(j_{~\alpha^0_n}+1)/2n \to\infty$ can occur only if
$j_{~\alpha^0_n}$ increases with $n$ ($a$ should not increase beyond $\beta\|\hat{u}\|_1$ because $(\beta\|\hat{u}\|_1)^a/a!$ is maximal roughly at this value.)
Then the quantity in the square bracket tends to $\pi\lambda_\beta^2 \left(\lambda_u^{-2}-c\right)$; therefore, if $n$ is large enough, $E_{\rm av}$ is a negative multiple of $\|~\alpha^0_n\|$ with a positive
$O(n)=o(\|~\alpha^0_n\|)$ correction coming from the entropy.

Consider now the random exponent
\bea\label{random-E}
E=
-\|~\alpha^0_n\|\left[\pi\lambda_\beta^2 \left(\frac{n}{\|~\alpha^0_n\|}\left[\overline{\left(X^0_{^\cdot}\right)^2}-\overline{X^0_{^\cdot}}^2\right]\right)
-\frac{2}{j_{~\alpha^0_n}+1}\left(\ln\frac{\beta\|\hat{u}\|_1}{a}+1-\frac{\ln 2\pi a}{2a}\right)\right]
+\ln {n-j_{~\alpha^0_n}\choose i_{~\alpha^0_n}}
\nonumber\\
=
-\pi\lambda_\beta^2 n\left[\overline{\left(X^0_{^\cdot}\right)^2}-\overline{X^0_{^\cdot}}^2\right]
+a\, i_{~\alpha^0_n} j_{~\alpha^0_n}\left(\ln\frac{\beta\|\hat{u}\|_1}{a}+1-\frac{\ln 2\pi a}{2a}\right)
+\ln {n-j_{~\alpha^0_n}\choose i_{~\alpha^0_n}}.
\eea
We assume that there is no concentration of probability at zero, i.e.
\be\label{no-concentration}
\Pp\left(\lim_{n\to\infty}\frac{n}{\|~\alpha_n\|}\left[\overline{\left(X^0_{^\cdot}\right)^2}-\overline{X^0_{^\cdot}}^2\right]>0  \right)=1,
\ee
reasonable because
\[
\Ee\left[\frac{n}{\|~\alpha_n\|}\left[\overline{\left(X^0_{^\cdot}\right)^2}-\overline{X^0_{^\cdot}}^2\right]\right]\to\lambda_u^{-2}-c>0.
\]
Then, due to $j_{~\alpha^0_n}\to\infty$, the
quantity
in the square bracket in the middle member of (\ref{random-E}) becomes positive and the random exponent also tends to minus infinity as $n$ increases.
Would (\ref{no-concentration}) fail, $~\alpha$ with $\|~\alpha_n\|/n\to\infty$ would still be
asymptotically
irrelevant. If in Eq.~(\ref{random-E}) $E<0$ or $E=o(n)$ positive, the contribution is negligible compared to that of other $~\alpha$'s yielding $E\propto n$, see below. $E$ cannot be positive and increase faster than $n$ because that would be incompatible with a stable pair potential.
It remains the possibility that $i_{~\alpha^0_n} j_{~\alpha^0_n}\sim n$, but then $i_{~\alpha^0_n}=o(n)$ and therefore the entropy is also $o(n)$. If $\overline{\left(X^0_{^\cdot}\right)^2}-\overline{X^0_{^\cdot}}^2$ could be of order one with a non-vanishing probability then for $\beta$ large enough $E$ would tend to minus infinity. At last, suppose that the only term of order $n$ is the middle one. This is independent of the variables $\{x^k_{j,r},\,x^{+k}_{j,r}\}$ and, hence, of the pair interactions. With the choice $a=\beta\|\hat{u}\|_1$ and counting with a similar contribution from the other cycles, a multiple of $\|\hat{u}\|_1=u(0)$ would be subtracted from the free energy per particle and, in effect, from the ground state energy per particle. This, however, is in contradiction with some earlier results [L], [S5] saying that for positive type pair potentials at high densities the distribution of the particles is uniform and the ground state energy per particle is $\rho\hat{u}(0)/2$. This is exactly the quantity in the exponential factor that appears in the partition function but is separated from $L^d F[n, \{n_l\}_1^p](0)$, see Eq.~(\ref{F(0)}). $\rho\hat{u}(0)/2$ is the mean-field result, at $\beta<\infty$ it overshoots the real value, and the negative correction to it should depend on the temperature. We conclude that {\em the contribution of all the $~\alpha_n$ whose norm diverges faster than $n$ becomes asymptotically negligible compared to that coming from $\|~\alpha_n\|=O(n)$}.

\vspace{3pt}
\noindent
(viii) On physical grounds we expect that the leading-order $e^{O(n)}$ factor by which a cycle of length $n$ contributes to the partition function is actually of order $e^{cn}$ with some $c=c(\beta)>0$. Before proving this let us recall that previously we have already met terms of order $e^{o(n)}$.
A rational $~\alpha$ leads to $n\left[\overline{\left(X^0_{^\cdot}\right)^2}-\overline{X^0_{^\cdot}}^2\right]=O(1)$ and at the same time the entropy is of order $\ln n$, so altogether we have a positive exponent of order $\ln n$ for such terms. Terms with $~\alpha$ irrational but $\|~\alpha_n\|=o(n)$ also produce an entropy-dominated factor $e^{o(n)}>1$. Let $g\geq 1$ be any monotone increasing function that tends to infinity and let $K\subset\Nn_+$ be a lacunary sequence satisfying
\[
\lim_{x\to \infty}\frac{\#\{k\in K|k\leq x\}}{x/g(x)}=1.
\]
Define
\[
\alpha^k_{k-j}=\left\{\begin{array}{ll}
a & {\rm if}\ j\leq j_0, k\in K\\
0 & {\rm otherwise}
\end{array}\right.
\]
Then
\be\label{relevant}
\|~\alpha^0_n\|\propto i_{~\alpha^0_n}\sim \frac{n}{g(n)},\quad \ln {n-j_{~\alpha^0_n}\choose i_{~\alpha^0_n}}\sim \frac{n}{g(n)}(\ln g(n)+1),
\ee
i.e. the entropy wins. All these terms become asymptotically independent of $\beta$ and will not appear in the limiting free energy density.

To show that the leading-order contribution is due to $\|~\alpha_n\|\propto n$ we apply the Markov inequality to $\frac{n}{\|~\alpha_n\|}\left[\overline{\left(X^0_{^\cdot}\right)^2}-\overline{X^0_{^\cdot}}^2\right]$. Choose some $A>1$, then
\[
\lim_{n\to\infty}\Pp\left(\frac{n}{\|~\alpha_n\|}\left[\overline{\left(X^0_{^\cdot}\right)^2}-\overline{X^0_{^\cdot}}^2\right]
< A (\lambda_u^{-2}-c) \right)\geq 1-\frac{1}{A}.
\]
Thus, for large enough systems
\be\label{exponent}
E
\gtrsim
\|~\alpha^0_n\|
\left[
-A \pi\lambda_\beta^2(\lambda_u^{-2}-c)
+ \frac{2}{j_{~\alpha^0_n}+1}\left(\ln\frac{\beta\|\hat{u}\|_1}{a}+1-\frac{\ln 2\pi a}{2a}\right) \right]
+\ln {n-j_{~\alpha^0_n}\choose i_{~\alpha^0_n}}
\ee
with a probability not smaller than $1-1/A$. We want $E$ to be positive.
A typical example for $\|~\alpha^0_n\|\propto n$ is (\ref{intracycle}) with $c=0$ and $Y^0_0$ tending to zero with probability one. However, for $\beta$ large enough the first term on the right side of (\ref{exponent}) is negative, to keep it smaller in modulus than the entropy the prefactor of $n$ in $\|~\alpha^0_n\|$ should decrease as $\beta$ increases. Let $i_{~\alpha^0_n}=\epsilon n$, then the entropy, computed with $j_{~\alpha^0_n}=O(1)$ is
\be
\ln{n-j_{~\alpha^0_n}\choose \epsilon n}\approx-n[\epsilon\ln\epsilon+(1-\epsilon)\ln(1-\epsilon)]
=
\epsilon n\ln \left[\epsilon^{-1}\left(1-\epsilon\right)^{-\frac{1-\epsilon}{\epsilon}} \right].
\ee
The best chance for the exponent (\ref{exponent}) to be positive is if $\|~\alpha^0_n\|$ is minimal under the condition that $i_{~\alpha^0_n}=\epsilon n$. The minimum is attained with $\alpha^k_{k-1}=1$ if $k\in K$ where $K\subset\{2,\dots,n\}$, $|K|=\epsilon n$, and $\alpha^k_{k-j}=0$ otherwise, resulting $\|~\alpha^0_n\|=\epsilon n$. The exponent is then
\be\label{energy+entropy}
\epsilon\left[ \ln \left(\epsilon^{-1}\left(1-\epsilon\right)^{-\frac{1-\epsilon}{\epsilon}} \right)
-A\pi(\lambda_\beta/\lambda_u)^2 +\ln\frac{e\beta\|\hat{u}\|_1}{\sqrt{2\pi}}  \right]n,
\ee
which is positive if
\be\label{bound-epsilon}
\epsilon(1-\epsilon)^{\frac{1-\epsilon}{\epsilon}}<e^{-A\pi(\lambda_\beta/\lambda_u)^2}.
\ee
Using $\frac{1-\epsilon}{\epsilon}\ln(1-\epsilon)=\sum_{k=1}^\infty\frac{\epsilon^k}{k(k+1)}-1$ it is seen that the left side of this inequality is a monotone increasing function of $\epsilon$ with the bounds
\[
\epsilon/e\leq \epsilon(1-\epsilon)^{\frac{1-\epsilon}{\epsilon}}\leq \epsilon.
\]
It follows that (\ref{bound-epsilon}) holds if $\epsilon<e^{-A\pi(\lambda_\beta/\lambda_u)^2}$ and fails if $\epsilon>e^{-A\pi(\lambda_\beta/\lambda_u)^2+1}$.

We mentioned already an exceptional occurrence of $\|~\alpha^0_n\|\sim n$ when  Eq.~(\ref{typical}) fails, the difference (\ref{variance}) is $O(1/n)$ while the separate terms are of order one. We return to it with the simplest example, $\alpha^n_1=1$ and $\alpha^k_{j}=0$ otherwise. Then $\|~\alpha^0_n\|=n-1$, $\Ee[(Y^0_0)^2]=\lambda_u^{-2} (n-1)^2/n^2$. Moreover, (keeping $~\alpha^+=0$) $\Ee\left[\,\overline{\left(X^0_{^\cdot}\right)^2}\,\right]=\lambda_u^{-2}(n-1)/n$ and thus $\Ee\left[\overline{\left(X^0_{^\cdot}\right)^2}-\overline{X^0_{^\cdot}}^2\right]=\lambda_u^{-2}(n-1)/n^2$. This is exactly the same value as the one we get if $\alpha^k_{k-1}=1$ for some $k\geq 2$ and $\alpha^{k'}_{k'-j'}=0$ otherwise.  A similar agreement can be found between any rational sequence and a properly chosen $~\alpha^0_n$ with a finite number of nonzero $\alpha^k_{k-j}$ where even the smallest $k$ may increase with $n$ (so $~\alpha^0_n\to~0$) but all the $j$ are of order 1. This is because $\|~\alpha^0_n\|$ does not depend on the value of $k$, only on the number of those for which $\alpha^k_{k-j}>0$.
The associated entropy is also the same as that of rational sequences, of the order of $\ln n$. Therefore either the exponent is negative or it is positive and of order $\ln n$ with the conclusion that the overall contribution is asymptotically negligible.

\vspace{4pt}
\noindent
(ix) Concerning $Y^+_0$ and $~\alpha^+_n$, a non-negligible contribution may come from here. As a typical example,
assume that $j_{~\alpha^+_{n}}$ particles of cycle 0 are coupled with $i_{~\alpha^+_{n}}$ particles of cycle 1, so the number of nonzero $\alpha^{+k}_j$ is $i_{~\alpha^+_{n}}j_{~\alpha^+_{n}}$.
Suppose also that $n_1$ tends to infinity. The same estimations can be made as for $~\alpha^0$. Keeping $j_{~\alpha^+_{n}}$ bounded, the largest contribution is for $i_{~\alpha^+_{n}}=\epsilon n_1$ with $\epsilon<e^{-(\pi/\eta)(\lambda_\beta/\lambda_u)^2}$.
However, computing the analogue of the exponent (\ref{random-E}) for cycle 1 we find it tending to $-\infty$. The reason is that
viewed from cycle 1 the roles of $j$ and $k$ are interchanged, $\alpha^{+k}_j>0$ for $\epsilon n_1$ different $k$ has the same effect on cycle 1
as $\alpha^{+k}_j>0$ for $\epsilon n$ different $j$ has on cycle 0: the sum to consider,
$\sum_{j=1}^n\sum_{k=1}^{n_1}k\,\alpha^{+k}_j$ diverges as $n_1^2$ while the maximum entropy associated with cycle 1 can increase only as $n_1$.
The conclusion is that $\|~\alpha^+_{n}\|$ must remain bounded.
{\em Thus, the leading contribution of cycle 0 to $L^d F^N_n(0)$ is a factor
$\sim \exp\left\{c e^{-(\pi/\eta) (\lambda_\beta/\lambda_u)^2}n \right\}$ with some $0<\eta<1$,
coming from $~\alpha=\{~\alpha^0, ~\alpha^+\}$, where $~\alpha^0$ is irrational with $\|~\alpha^0_n\| \propto e^{-(\pi/\eta)(\lambda_\beta/\lambda_u)^2}n$ and $~\alpha^+$ is rational, i.e. $\|~\alpha^+\|<\infty$}. For the exponential increase in $n$ the non-decay of $L^d f_n(0;L\overline{X^0_{^\cdot}})$ is also necessary. If $n\lambda_\beta^2/L^2$ is bounded, this holds true irrespective of the value of $\overline{X^0_{^\cdot}}$, cf. (\ref{f_n(0)-asymp}); if $n\lambda_\beta^2/L^2$ tends to infinity, this is a consequence of $|\overline{X^0_{^\cdot}}|=O(1/\sqrt{n})$ to be shown below.

Although the role of the coupling of any individual cycle of a diverging length to the other cycles appears to be negligible, for the whole system the coupling among cycles results in a decrease of the free energy density. This is because for $\beta<\infty$ the partitions of $N$ to $p\propto N$ elements dominate the partition function: single particles and particles forming finite cycles carry the uncondensed density which is positive at positive temperatures.
An example in [S3, Remark 6] has shown that for any nonzero value of $\beta\rho\hat{u}(0)$ the dominant contribution to the partition function comes from $p\propto N$ and $~\alpha$ chosen so that  in the graph ${\cal G}_{~\alpha}$ a macroscopic number of cycles occur in couplings.

\vspace{5pt}
\noindent
(x) In all the cases of $\|~\alpha_n\|/n=O(1)$ when $~\alpha^0_n$ is the restriction of an infinite $~\alpha^0$ to $\{\alpha^k_{k-j}|1\leq j<k\leq n\}$ and $~\alpha^+$ is rational $\overline{X^0_{^\cdot}}\to 0$ almost surely. To establish the rate of decay
the simplest result is obtained by using Chebyshev's inequality: for any $A>1$
\be
\Pp\left(|\overline{X^0_{^\cdot}}|< A \sqrt{\Ee\left[\overline{X^0_{^\cdot}}^2\right]}  \right) >1-1/A^2.
\ee
A bound on the standard deviation can be inferred from a comparison of Eqs.~(\ref{alpha-norms}) and (\ref{EY0-EY+}). Setting $\alpha^k_{k-j}=0$ for $j>j_{~\alpha^0_n}$ and $\alpha^{+k}_j=0$ for $j>j_{~\alpha^+_n}$ with $j_{~\alpha^0_n},\, j_{~\alpha^+_n}=O(1)$ justified by the previous discussion, we find
\be
\sqrt{\Ee\left[\overline{X^0_{^\cdot}}^2\right]}
\leq
\frac{1}{n\lambda_u}\sqrt{j_{~\alpha^0_n}\|~\alpha^0_n\|+j_{~\alpha^+_n}\|~\alpha^+_n\|}
\leq \frac{c_1 e^{-c_2(\lambda_\beta/\lambda_u)^2}}{\lambda_u\sqrt{n}}.
\ee
Sharper results can be obtained by using Bernstein's inequality for compactly supported $\hat{u}$ or by making other assumptions about its fast decay.

\vspace{5pt}
\noindent
(xi) The sign oscillation seen in $F\left[n,\{n_l\}_1^p\right](0)$ is to a large extent due to the stability of the interaction. In the Fourier-representation we use here instability shows up in $\int \hat{u}(x)\d x=u(0)<0$. In the extreme case when $\hat{u}\leq 0$ all the terms of $F\left[n,\{n_l\}_1^p\right](0)$ are positive. The positive-type $u$ considered in this paper represents the opposite extremity: $\hat{u}\geq 0$ defines a stable interaction, and the signs alternate according to the parity of the sum  $\sum_{j<k}\alpha^k_j$. In the finite-volume canonical ensemble the difference between stable and unstable interactions is not striking. Stability is in the background, it guaranties the existence of the thermodynamic limit of the free energy density and, more generally, of the Gibbs measure, missing in unstable systems. $F\left[n,\{n_l\}_1^p\right](0)>0$, so the negative terms cannot flip the overall sign. Moreover, any accidental cancellation would only make the almost sure convergence of $\overline{X^0_{^\cdot}}$ to zero faster, and therefore would not jeopardize the proof of the second part of the theorem. However, because the variables occurring in terms of different signs are partly independent, exact cancellation between positive and negative terms is an event of zero probability, and therefore the order-of-magnitude estimates are not altered by them.

To summarize, $\nu^{N,L}_n(y)$ is asymptotically concentrated to an $O(1/\sqrt{n})$ neighborhood of the origin.
This outcome is physically satisfying: if and only if $\|~\alpha_n\|\propto n$ (and then $|\overline{X^0_{^\cdot}}|\propto 1/\sqrt{n}$)  the contributions of cycle 0 to the energy and the entropy are of the same order, and this is the only case when the temperature has an influence on what the cycle adds to the free energy density.
The implication of the concentration of $\nu^{N,L}_n$ to $\delta_0$ is that
\[
\lim_{n\to\infty} F^{{\cal N}(n)}_n(x)\, /F^{{\cal N}(n)}_n(0)=1.
\]
It then follows that $\sigma_1(x)\geq \rho-\sum_{n=1}^\infty \rho_n$. For an upper bound we add to the lower one the contribution (\ref{sigma1-from-finite n}) of all the finite $n$ and end up with the result (\ref{cond1}).

\subsection{Condition for BEC}

Here we do not use the asymptotic form of $f(x;\overline{Z^0_{^\cdot}})$, because the exact expressions are directly obtained, see Eq.~(\ref{sum-vs-first_term}):
\bea\label{L-dfn(0)}
L^d f_n\left(0;L\overline{X^0_{^\cdot}}\right)
&=&
\sum_{z\in\Zz^d}\exp\left\{-\frac{\pi n \lambda_\beta^2}{L^2}\left(z+ L\overline{X^0_{^\cdot}}\right)^2\right\}
\nonumber\\
&=&
\exp\left\{-\pi n \lambda_\beta^2 \overline{X^0_{^\cdot}}^2\right\}
\sum_{z\in\Zz^d}\exp\left\{-\frac{\pi n \lambda_\beta^2}{L^2}\left(z^2+2z\cdot L\overline{X^0_{^\cdot}}\right)\right\}
\eea
and
\be\label{int-f}
\int_\Lambda f_n\left(x;L\overline{X^0_{^\cdot}}\right)\d x
=\exp\left\{-\pi n \lambda_\beta^2 \overline{X^0_{^\cdot}}^2\right\}
=\frac{L^d f_n\left(0;L\overline{X^0_{^\cdot}}\right)}{\sum_{z\in\Zz^d}\exp\left\{-\frac{\pi n \lambda_\beta^2}{L^2}\left(z^2+2z\cdot L\overline{X^0_{^\cdot}}\right)\right\}}.
\ee
Therefore
\bea\label{intFn(x)-asymp}
\int_\Lambda F\left[n_0,\{n_l\}_1^p\right](x)\d x
=  \sum_{\alpha^2_1,\alpha^3_1,\alpha^3_2,\dots,\alpha^N_{N-1}=0}^\infty\ \left(L^{-d}\right)^{K_{\{\alpha^k_j\}}}
\prod_{1\leq j<k\leq N}
\frac{\left(-\beta\right)^{\alpha^k_j}}{\alpha^k_j !} \prod_{r=1}^{\alpha^k_j}
\int\d x^k_{j,r}\ \hat{u}\left(x^k_{j,r}\right) \int_0^1\d t^k_{j,r}
\nonumber\\
\left[
\delta(X^0_1,\dots,X^p_1) \prod_{l=0}^p
\exp\left\{-\pi n_l \lambda_\beta^2\left[\overline{\left(X^l_{^\cdot}\right)^2}-\overline{X^l_{^\cdot}}^2\right]\right\}
\sum_{z\in\Zz^d}\exp\left\{-\frac{\pi n_l \lambda_\beta^2}{L^2}\left(z+L\overline{X^l_{^\cdot}}\right)^2\right\}
\right.
\nonumber\\
\left.
\times\frac{1}{\sum_{z\in\Zz^d}\exp\left\{-\frac{\pi n_0 \lambda_\beta^2}{L^2}\left(z^2+2z\cdot L\overline{X^0_{^\cdot}}\right)\right\}}
\right]\,.
\nonumber\\
\eea
Without the last fraction the above expression is $L^d F\left[n_0,\{n_l\}_1^p\right](0)$. Thus, it is the asymptotic behavior of this fraction that decides about BEC. If $n_0=o(N^{2/d})$, the denominator tends to infinity irrespective of the value of $\overline{X^0_{^\cdot}}$: if $L\overline{X^0_{^\cdot}}$ remains bounded or increases slower than $N^{2/d}/n_0$ then each term tends to 1; otherwise infinitely many terms with $z\cdot L\overline{X^0_{^\cdot}}<0$ will exceed 1. As a result, such cycles do not add to the condensate. Therefore, we focus on cycles of length $n\geq cN^{2/d}$ where $c>0$.
With the definitions (\ref{HN-n}) and (\ref{Theta})
\bea
\lefteqn{
\int_\Lambda F^N_n(x)\d x
=
\sum_{\{\alpha^k_j\in\Nn_0|1\leq j\leq n,\, j+1\leq k\leq N\}} \left(\prod_{ j=1}^n\prod_{k=j+1}^N\frac{\left(-\beta\right)^{\alpha^k_j}}{\alpha^k_j !}\right)
\int_0^1 \prod_{ j=1}^n\prod_{k=j+1}^N\prod_{r=1}^{\alpha^k_j}\d t^k_{j,r}     }
\nonumber\\
&&\int \prod_{j=1}^n\prod_{k=j+1}^N \prod_{r=1}^{\alpha^k_j}\d x^k_{j,r}\hat{u}\left(x^k_{j,r}\right)
H_{N-n}\
\frac{\exp\left\{-\pi n \lambda_\beta^2\left[\overline{\left(X^0_{^\cdot}\right)^2}-\overline{X^0_{^\cdot}}^2\right]\right\}
L^d f_n\left(0;L\overline{X^0_{^\cdot}}\right)}{\sum_{z\in\Zz^d}\exp\left\{-\frac{\pi n \lambda_\beta^2}{L^2}\left(z^2+2z\cdot L\overline{X^0_{^\cdot}}\right)\right\}}
\nonumber\\
&&=
\int_{\Rr^d}\d y \frac{\Theta^N_n(y)}{\sum_{z\in\Zz^d}\exp\left\{-\frac{\pi n_0 \lambda_\beta^2}{L^2}\left(z^2+2z\cdot Ly\right)\right\}}.
\eea
Dividing by $L^d F^N_n(0)=\int\Theta^N_n(y')\d y'$
\be
\rho^{N,L}_0
\geq
\sum_{n\geq cN^{2/d}}\rho^{N,L}_n\, \frac{\int_\Lambda F^N_n(x)\d x}{L^d F^N_n(0)}
=
\sum_{n\geq cN^{2/d}}\rho^{N,L}_n\,\int_{\Rr^d} \frac{\nu^{N,L}_n(y)\,\d y}{\sum_{z\in\Zz^d}\exp\left\{-\frac{\pi n \lambda_\beta^2}{L^2}z\cdot\left(z+2Ly\right)\right\}}.
\ee
The result (\ref{cond2}) for BEC follows from the fact that asymptotically $\nu^{N,L}_n$ is concentrated to an $O(1/\sqrt{n})$ neighborhood of the origin. For $n\propto N^{2/d}$ this means that $L|y|=O(1)$, and if $n/N^{2/d}\to\infty$ then $L|y|=o(1)$; thus, for any $n\geq cN^{2/d}$ the sum over $\Zz^d$ remains finite in the thermodynamic limit.

This ends the proof for pair potentials with an existing $\hat{u}\in L^1\cap C(\Rr^d)$ of compact support. The latter condition can be dropped preserving $\int|\hat{u}(x)|x^2\d x<\infty$, with the only effect that the sure convergence to zero is changed into almost sure one.

\vspace{20pt}
\noindent{\Large\bf References}
\begin{enumerate}
\item[{[ACK]}] Adams S., Collevecchio A., and K\"onig W.: {\em A variational formula for the free energy of an interacting many-particle system.} Ann. Prob. {\bf 39}, 683-728 (2011).
\item[{[AD]}] Adams S. and Dickson M.: {\em An explicit large deviations analysis of the spatial cycle Huang-Yang-Luttinger model.} Ann. Henri Poincar\'e {\bf 22}, 1535-1560 (2021).
\item[{[AL]}] Aizenman M. and Lieb E. H.: {\em Magnetic properties of some itinerant-electron systems at $T>0$.} Phys. Rev. Lett. {\bf 65}, 1470-1473 (1990).
\item[{[AN]}] Aizenman M. and Nachtergaele B.: {\em Geometric aspects of quantum spin states.} Commun. Math. Phys. {\bf 164}, 17-63 (1994).
\item[{[BCMP]}] Benfatto G., Cassandro M., Merola I. and Presutti E.: {\em Limit theorems for statistics of combinatorial partitions with applications to mean field Bose gas.} J. Math. Phys. {\bf 46}, 033303 (2005).
\item[{[BP]}] Buffet E. and Pul\'e J. V.: {\em Fluctuation properties of the imperfect Bose gas.} J. Math. Phys. {\bf 24}, 1608-1616 (1983).
\item[{[BSch]}] Bund S. and Schakel M. J.: {\em String picture of Bose-Einstein condensation.} Mod. Phys. Lett. B {\bf 13}, 349 (1999).
\item[{[BU1]}] Betz V. and Ueltschi D.: {\em Spatial random permutations and infinite cycles.} Commun. Math. Phys. {\bf 285}, 469-501 (2009).
\item[{[BU2]}] Betz V. and Ueltschi D.: {\em Spatial random permutations and Poisson-Dirichlet law of cycle lengths.} Electr. J. Probab. {\bf 16}, 1173-1192 (2011).
\item[{[BUV]}] Betz V., Ueltschi D. and Velenik Y.: {\em Random permutations with cycle weights.} Ann. Appl. Probab. {\bf 21}, 312-331 (2011).
\item[{[Ce]}] Ceperley D. M.: {\em Path integrals in the theory of condensed helium.} Rev. Mod. Phys. {\bf 67}, 279-355 (1995).
\item[{[DMP]}] Dorlas T. C., Martin Ph. A. and Pul\'e J. V.: {\em Long cycles in a perturbed mean field model of a boson gas.} J. Stat. Phys. {\bf 121}, 433-461 (2005).
\item[{[EP]}] Elboim D. and Peled R.: {\em Limit distributions for Euclidean random permutations.} Commun. Math. Phys. {\bf 369}, 457-522 (2019).
\item[{[F1]}] Feynman R. P.: {\em Atomic theory of the $\lambda$ transition in helium.} Phys. Rev. {\bf 91}, 1291-1301 (1953).
\item[{[F2]}] Feynman R. P.: {\em Space-time approach to non-relativistic quantum mechanics.} Rev. Mod. Phys. {\bf 20}, 367-387 (1948).
\item[{[Fe]}] Feller W.: {\em An introduction to probability theory and its applications. Vol. 2, Ch. VII.8. Theorem 3.} John Wiley \& Sons, New York (1970).
\item[{[G1]}] Ginibre J.: {\em Some applications of functional integration in Statistical Mechanics.} In: {\em Statistical Mechanics and Quantum Field Theory}, eds. C. De Witt and R. Stora, Gordon and Breach (New York 1971).
\item[{[G2]}] Ginibre J.: {\em Reduced density matrices of quantum gases. I. Limit of infinite volume.} J. Math. Phys. {\bf 6}, 238-251 (1965).
\item[{[G3]}] Ginibre J.: {\em Reduced density matrices of quantum gases. II. Cluster property.} J. Math. Phys. {\bf 6}, 252-262 (1965).
\item[{[G4]}] Ginibre J.: {\em Reduced density matrices of quantum gases. III. Hard-core potentials.} J. Math. Phys. {\bf 6}, 1432-1446 (1965).
\item[{[K1]}] Kac M.: {\em On distributions of certain Wiener functionals.} Trans. Amer. Math. Soc. {\bf 65}, 1-13 (1949).
\item[{[K2]}] Kac M.: {\em On some connections between probability theory and differential and integral equations.} In: Proceedings of the Second Berkeley Symposium on Probability and Statistics, J. Neyman ed., Berkeley, University of California Press (1951)
\item[{[L]}]  Lieb E. H.: {\em  Simplified approach to the ground-state energy of an imperfect Bose gas.} Phys. Rev. {\bf 130}, 2518-2528 (1963).
\item[{[LZP]}] Lewis J. T., Zagrebnov V. A. and Pul\'e J. V.: {\em The large deviation principle for the Kac distribution.} Helv. Phys. Acta {\bf 61} 1063-1078 (1988).
\item[{[PO]}] Penrose O. and Onsager L.: {\em Bose-Einstein condensation and liquid He.} Phys. Rev. {\bf 104}, 576-584 (1956).
\item[{[R]}] Ruelle D.: \emph{Statistical Mechanics.} W. A. Benjamin (New York-Amsterdam 1969).
\item[{[S1]}] S\"ut\H o A.: {\em Percolation transition in the Bose gas.} J. Phys. A: Math. Gen. {\bf 26}, 4689-4710 (1993).
\item[{[S2]}] S\"ut\H o A.: {\em Percolation transition in the Bose gas: II.} J. Phys. A: Math. Gen. {\bf 35}, 6995-7002 (2002). See also arXiv:cond-mat/0204430v4 with addenda after Eqs.~(34) and (44).
\item[{[S3]}] S\"ut\H o A.:  {\em Fourier formula for quantum partition functions.} arXiv:2106.10032 [math-ph] (2021).
\item[{[S4]}] S\"ut\H o A.: {\em Correlation inequalities for noninteracting Bose gases.} J. Phys. A: Math. Gen. {\bf 37}, 615-621 (2004).
\item[{[S5]}] S\"ut\H o A.: {\em Ground state at high density.} Commun. Math. Phys. {\bf 305}, 657-710 (2011).
\item[{[S6]}] S\"ut\H o A.: {\em The total momentum of quantum fluids.} J. Math. Phys. {\bf 56}, 081901 (2015), Section IV.
\item[{[Sch]}] Schakel A. M. J.: {\em Percolation, Bose-Einstein condensation, and string proliferation.} Phys. Rev. E {\bf 63}, 026115 (2001).
\item[{[T]}] T\'oth B.: {\em Phase transition in an interacting Bose system. An application of the theory of
Ventsel’ and Freidlin.} J. Stat. Phys {\bf 61}, 749–764 (1990).
\item[{[U1]}] Ueltschi D.: {\em Relation between Feynman cycles and off-diagonal long-range order.} Phys. Rev. Lett. {\bf 97}, 170601 (2006).
\item[{[U2]}] Ueltschi D.: {\em Feynman cycles in the Bose gas.} J. Math. Phys. {\bf 47}, 123303 (2006).
\item[{[V]}] Vershik A. M.: {\em Statistical mechanics of combinatorial partitions, and their limit shapes.} Funk. Anal. Appl. {\bf 30}, 90-105 (1996).
\item[{[Y]}] Yang C. N.: {\em Concept of off-diagonal long-range order and the quantum phases of liquid He and of superconductors.} Rev. Mod. Phys. {\bf 34}, 694-704 (1962).
\end{enumerate}

\end{document}